\documentclass[epj,nopacs]{svjour}

\usepackage{graphicx}
\usepackage{latexsym}
\usepackage{amsmath}
\usepackage{amssymb}
\usepackage{amsfonts}
\usepackage{bm}
\usepackage{color}

\newcommand{\la}{\left<}
\newcommand{\ra}{\right>}
\newcommand{\nvec}{\underline{n}}

\newcommand{\qvec}{\ensuremath{{\bf q}}}
\newcommand{\rvec}{\ensuremath{{\bf r}}}
\newcommand{\bfzero}{\ensuremath{{\bf 0}}}
\newcommand{\ddiff}{\ensuremath{\mathrm{d}}}

\newcommand{\muAc}{\ensuremath{\mu_{\mathrm{A},c}}}
\newcommand{\muFc}{\ensuremath{\mu_{\mathrm{F},c}}}

\newcommand{\tincr}{\delta \tau}
\newcommand{\tsamp}{\Delta \tau}
\newcommand{\tsampmax}{{\Delta \tau}_{\mathrm{max}}}

\newcommand{\xbf}{\ensuremath{\mathbf{x}}}

\newcommand{\Nc}{\ensuremath{n_\mathrm{c}}}

\newcommand{\Nk}{\ensuremath{n_\mathrm{k}}}
\newcommand{\Nl}{\ensuremath{n_l}}
\newcommand{\Nt}{\ensuremath{n_\mathrm{t}}}
\newcommand{\Nr}{\ensuremath{n_{\rvec}}}
\newcommand{\Fcal}{\ensuremath{{\cal F}}}
\newcommand{\Ocal}{\ensuremath{{\cal O}}}
\newcommand{\Aop}{\ensuremath{\mathbf{A}}}
\newcommand{\Bop}{\ensuremath{\mathbf{B}}}
\newcommand{\Eop}{\ensuremath{\mathbf{E}}}
\newcommand{\Pop}{\ensuremath{\mathbf{P}}}
\newcommand{\Vop}{\ensuremath{\mathbf{V}}}
\newcommand{\Sop}{\ensuremath{\mathbf{\cal L}_{\tsamp}}}

\newcommand{\dOtot}{\ensuremath{\delta \Ocal^2_{\mathrm{tot}}}}
\newcommand{\dOint}{\ensuremath{\delta \Ocal^2_{\mathrm{int}}}}
\newcommand{\dOext}{\ensuremath{\delta \Ocal^2_{\mathrm{ext}}}}
\newcommand{\sOtot}{\ensuremath{\delta \Ocal_{\mathrm{tot}}}}
\newcommand{\sOint}{\ensuremath{\delta \Ocal_{\mathrm{int}}}}
\newcommand{\sOext}{\ensuremath{\delta \Ocal_{\mathrm{ext}}}}
\newcommand{\dvtot}{\ensuremath{\delta v^2_{\mathrm{tot}}}}
\newcommand{\svtot}{\ensuremath{\delta v_{\mathrm{tot}}}}
\newcommand{\dvint}{\ensuremath{\delta v^2_{\mathrm{int}}}}
\newcommand{\svint}{\ensuremath{\delta v_{\mathrm{int}}}}
\newcommand{\dvext}{\ensuremath{\delta v^2_{\mathrm{ext}}}}
\newcommand{\svext}{\ensuremath{\delta v_{\mathrm{ext}}}}
\newcommand{\dvgauss}{\ensuremath{\delta v^2_{\mathrm{G}}}}
\newcommand{\svgauss}{\ensuremath{\delta v_{\mathrm{G}}}}

\newcommand{\dVcell}{\delta V}

\newcommand{\Dnonerg}{\Delta_{\mathrm{ne}}^2}
\newcommand{\Snonerg}{\Delta_{\mathrm{ne}}}
\newcommand{\Tnonerg}{\tau_{\mathrm{ne}}}

\newcommand{\Nnonerg}{N_{\mathrm{ne}}}
\newcommand{\Tglass}{T_{\mathrm{g}}}
\newcommand{\taubasin}{\tau_{\mathrm{b}}}
\newcommand{\ttemper}{\tau_{\mathrm{temp}}}
\newcommand{\taualph}{\tau_{\alpha}}

\newcommand{\gamint}{\gamma_{\mathrm{int}}}
\newcommand{\gamext}{\gamma_{\mathrm{ext}}}
\newcommand{\gaminthat}{\hat{\gamma}_{\mathrm{int}}}
\newcommand{\gamexthat}{\hat{\gamma}_{\mathrm{ext}}}

\newcommand{\gamextapp}{\tilde{\gamma}_{\mathrm{ext}}}

\newcommand{\sr}{s_{\rvec}}
\newcommand{\scr}{s_{c\rvec}}
\newcommand{\fr}{g_{\rvec}}
\newcommand{\fcr}{g_{c\rvec}}

\newcommand{\daD}{d_\mathrm{1}}

\begin{document}

\title{Simple models for strictly non-ergodic stochastic processes of macroscopic systems}

\author{G. George
\and L. Klochko
\and A.N. Semenov
\and J. Baschnagel
\and J.P.~Wittmer\thanks{joachim.wittmer@ics-cnrs.unistra.fr}
}
\institute{Institut Charles Sadron, Universit\'e de Strasbourg \& CNRS, 23 rue du Loess, 67034 Strasbourg Cedex, France}
\date{Received: date / Revised version: date}

\abstract{We investigate simple models for strictly non-ergodic stochastic processes $x_t$ 
($t$ being the discrete time step) focusing on the expectation value $v$ and the standard 
deviation $\delta v$ of the empirical variance $v[\xbf]$ of finite time series $\xbf$. 
$x_t$ is averaged over a fluctuating field $\sigma_{\rvec}$ ($\rvec$ being the microcell 
position) characterized by a quenched spatially correlated Gaussian field $\fr$.
Due to the quenched $\fr$-field $\delta v(\tsamp)$ becomes 
a finite constant, $\Snonerg > 0$, for large sampling times $\tsamp$.
The volume dependence of the non-ergodicity parameter $\Snonerg$ is 
investigated for different spatial correlations.
Models with marginally long-ranged $\fr$-correlations are successfully mapped on 
shear-stress data from simulated amorphous glasses of polydisperse beads.}
\maketitle

\section{Introduction}
\label{sec_intro}

It is common to characterize a stochastic process $x(\tau)$ 
\cite{VanKampenBook} using ensembles $\{\xbf\}$ of discrete time series
\begin{equation}
\xbf = \{x_t=x(\tau_t = t \tincr),t=1,\ldots,\Nt\}
\label{eq_xbf_def}
\end{equation}
with $\tau$ being the continuous time, $t$ the discrete time,
$\tincr$ the time interval between the equidistant measurements
and $\tsamp = \Nt \tincr$ the experimentally or computationally available
``sampling time" \cite{WXP13,WKC16,lyuda19a,spmP1,spmP2}. 
Let us denote by $\Ocal[\xbf]$ a functional of a given time series $\xbf$.
If the stochastic process $x(\tau)$ is {\em ergodic} \cite{PlischkeBook},
the expectation value $\Ocal$ and the standard deviation $\delta \Ocal$ of $\Ocal[\xbf]$ 
may be obtained by either averaging over ensembles $\{\xbf_c, c = 1,\ldots, \Nc \}$ 
of independent ``configurations" $c$ (``$c$-averaging")
or over ensembles $\{\xbf_k, k = 1,\ldots, \Nk \}$ of time series $k$ 
of one large trajectory $c$ (``$k$-averaging")
exploring a significant representative part of the generalized phase space of the system. 
It is thus sufficient for such ergodic systems to 
characterize the time series $\xbf$ by {\em one} 
index $c$ or $k$.\footnote{We assume $\Nc\gg 1$ and $\Nk \gg 1$ throughout this work.}

The ergodicity hypothesis is in fact violated in many physical, biological and 
socio-economic systems, i.e. even very long ``$c$-trajectories" remain trapped 
(at least in practice) 
in ``meta-basins" of a generalized phase space 
\cite{VanKampenBook,spmP1,spmP2,PlischkeBook,Heuer08,Gardner}.
(For Hamiltonian dynamical systems such basins correspond simply to valleys of the 
potential energy landscape \cite{Heuer08}, 
for more general stochastical dynamical schemes to valleys of the relevant {\em free} energy landscape
quantified by the minimal external work needed to quasistatically push the system into a specific state point.)
Modelling the statistics and dynamics of such {\em non-ergodic} processes has become of paramount 
importance, especially in conjunction with advanced experimental techniques, such as single particle 
tracing in cells \cite{Metzler14}.
Importantly, a time series $\xbf_{ck}$ must now be characterized by {\em two} indices $c$ and $k$ 
and it becomes crucial in which order $c$- and $k$-averages are taken \cite{spmP2}.
As a consequence, the standard total variance 
\begin{equation}
\dOtot(\tsamp) = \dOint(\tsamp) + \dOext(\tsamp)
\label{eq_intro_dOtot}
\end{equation}
is the sum of {\em two} contributions characterizing, respectively, 
the internal variance within each $c$ and the external variance between 
different $c$. Moreover, for large sampling times $\sOint \to 0$
while $\delta \Ocal(\tsamp) \simeq \sOext(\tsamp)$ approaches for non-ergodic systems 
a positive definite constant $\Snonerg \equiv \lim_{\tsamp \to \infty} \sOext(\tsamp)$.
This is the relevant ``non-ergo\-di\-city parameter" \cite{spmP1,spmP2} of this study.
(See Sec.~\ref{theo_ck} for more details.)
Fortunately, $\Snonerg$ decreases generally with the system volume $V$
for processes with a large number $\Nr \propto V$ of more or less independent microcells \cite{spmP1}. 

One goal of the present work is to introduce some useful
operator notations allowing to characterize concisely fluctuations of ensembles of non-ergodic systems and
to illustrate the above statements by means of 
various simple stochastic models which can be treated (essentially) analytically.
Moreover, we attempt to describe the system-size 
dependence of $\Snonerg$ by means of two-point spatial correlation functions 
of an effective quenched microscopic field $\fr$ related to
the $k$-averaged standard deviation $\sr$ of a microscopic fluctuating field 
$\sigma_{\rvec}$ ($\rvec$ labeling the microcell position).
As in our recent studies \cite{WKC16,lyuda19a,spmP1,spmP2}  
we focus on the empirical variance $v[\xbf]$ (defined in Sec.~\ref{theo_stationary})
and the corresponding expectation value $v(\tsamp)$ and the standard deviations $\svint(\tsamp)$ and $\svext(\tsamp)$.
One important motivation is that many physical quantities can be obtained by 
{\em equilibrium} molecular dynamics (MD) or Monte Carlo (MC) simulations 
\cite{AllenTildesleyBook,LandauBinderBook} 
using fluctuation dissipation relations 
\cite{PlischkeBook,AllenTildesleyBook,ChaikinBook,Lebowitz67,Lutsko89,Procaccia16}. 
Understanding how the respective variances and their 
standard deviations depend on the length $\tsamp$ of the production runs 
and the simulation box volume $V$ is thus crucial \cite{WKC16,lyuda19a,spmP1,spmP2,Procaccia16}.

We recall first in Sec.~\ref{sec_theo} recent results \cite{lyuda19a,spmP1,spmP2}
and discuss then in Sec.~\ref{sec_V} the $V$-dependence of various properties and, 
more specifically, how $\Snonerg(V)$ may depend on spatial correlations 
(Sec.~\ref{V_correl}) under the physically motivated constraint that the expectation value of the 
variance $v$ must be $V$-independent (Sec.~\ref{V_T}).
We turn then in Sec.~\ref{sec_spm} to the description 
of different imposed $\fr$-distribu\-tions (Sec.~\ref{spm_models}). 
Model variants are mapped in Sec.~\ref{sec_simu} onto simulated data obtained from the 
shear stresses in amorphous glasses 
\cite{WXP13,lyuda19a,spmP1,spmP2}.
Our results are summarized in Sec.~\ref{sec_conc}.
The numerical generation of spatially correlated Gaussian fields 
is discussed in Appendix~\ref{app_gaussfield} 
and an alternative quenched field 
important for future work \cite{spmP4} in Appendix~\ref{app_X2}.

\section{Makroscopic properties}
\label{sec_theo}

\subsection{Some useful notations}
\label{theo_definitions}

It is useful to introduce a few notations. The $l$-average operator 
\begin{equation}
\Eop^l \Ocal_{lmn\ldots}  
\equiv \frac{1}{\Nl} \sum_{l=1}^{\Nl} \Ocal_{lmn\ldots} \equiv \Ocal_{mn\ldots}(\Nl) \label{eq_Eopdef}
\end{equation}
takes a property $\Ocal_{lmn\ldots}$ depending possibly on several indices $l,m,\ldots$
and projects out the specified index $l$, 
i.e. the $l$-average $\Ocal_{mn\ldots}(\Nl)$ does not depend any more on $l$,
but it may depend on the upper bound $\Nl$ as marked by the argument.
Introducing the power-law operator $\Pop^{\alpha} \Ocal \equiv \Ocal^{\alpha}$,
with the exponent $\alpha=2$ being here the only relevant case,
and using the standard commutator $[\Aop,\Bop] \equiv \Aop \Bop - \Bop \Aop$ 
for two operators $\Aop$ and $\Bop$,
the $l$-variance operator is defined by $\Vop^l \equiv [\Eop^l,\Pop^2]$. 
Note that the $l$-variance
\begin{equation}
\delta \Ocal_{mn\ldots}^2(\Nl) \equiv \Vop^l \Ocal_{lmn\ldots}
\label{eq_Vopdef}
\end{equation}
depends as well in general on the upper bound $\Nl$.
For many cases considered below $\Ocal_{lmn\ldots}(\Nl)$ and 
$\delta \Ocal_{lmn\ldots}(\Nl)$ converge for large $\Nl$ (formally $\Nl \to \infty$)
or become stationary for the experimentally and numerically accessible $\Nl$-range.
This limit is denoted by $\Ocal_{mn\ldots}$ and $\delta \Ocal_{lmn\ldots}$
without the argument $\Nl$.
As discussed in detail in Ref.~\cite{spmP2}, 
we have defined $\Vop^l$ as an uncorrected (biased) 
sample variance operator without the standard Bessel correction \cite{LandauBinderBook}, 
i.e. we normalize with $1/\Nl$ and not with $1/(\Nl-1)$.
This difference is irrelevant for all cases with $\Nl \gg 1$.

\subsection{Extended $ck$-ensemble for non-ergodic systems}
\label{theo_ck}

As stated in the Introduction, for non-ergodic systems a time series $\xbf_{ck}$ 
must be characterized by {\em two} discrete indices $c$ and $k$ with $c$ standing 
for the independently generated configuration and $k$ for a subset of length $\Nt$ of a much larger 
trajectory generated for a fixed configuration $c$.
Importantly, the $k$-averages
\begin{eqnarray}
\Ocal_c(\tsamp,\Nk) & \equiv & \Eop^k \Ocal[\xbf_{ck}]
\mbox{ and } \label{eq_Ocal_c_def} \\
\delta \Ocal^2_c(\tsamp,\Nk)  & \equiv & \Vop^k \Ocal[\xbf_{ck}]
\label{eq_dOcal_c_def}
\end{eqnarray}
depend in general not only on the sampling time $\tsamp = \Nt \tincr$
and the number $\Nk$ of time series probed but also on $c$ 
(as marked by the index).\footnote{We 
assume in the present work that the longest relaxation time $\taualph$ of the system
becomes arbitrarily large, i.e. especially $\tsamp \ll\taualph$. 
The $c$-dependence drops out for ergodic systems with finite $\taualph$ and $\tsamp \gg \taualph$.
See Sec. 2.2.9 of Ref.~\cite{spmP2} for the $\Nc$-, $\Nk$- and $\tsamp$-dependences
in the latter limit.}
The three types of variances mentioned in Sec.~\ref{sec_intro} are defined by 
\begin{eqnarray}
\dOtot(\tsamp) & \equiv & [\Eop^c \Eop^k,\Pop^2] \Ocal[\xbf_{ck}] \label{eq_dOtot} \\
\dOint(\tsamp) & \equiv & \Eop^c \delta \Ocal_c^2(\tsamp) = \Eop^c \Vop^k \Ocal[\xbf_{ck}] \label{eq_dOint} \\
\dOext(\tsamp) & \equiv & \Vop^c \Ocal_c(\tsamp) = \Vop^c \Eop^k \Ocal[\xbf_{ck}].    \label{eq_dOext}
\end{eqnarray}
Using the identity $[\Eop^c \Eop^k,\Pop^2] = \Eop^c \Vop^k + \Vop^c \Eop^k$ \cite{spmP2},
it is seen that Eq.~(\ref{eq_intro_dOtot}) exactly holds.
The dependencies of the variances on $\tsamp$, $\Nc$ and $\Nk$ are discussed in detail in Ref.~\cite{spmP2}.
Importantly, the expectation value of $\sOtot(\tsamp)$ for $\Nc \to \infty$ 
is strictly $\Nk$-independent and may also be computed using $\Nk=1$.
$\dOtot(\tsamp)$ is thus the standard commonly computed variance 
\cite{Procaccia16,WKC16,lyuda19a,spmP1}.
The ``internal variance" $\dOint(\tsamp)$ and the ``external variance" $\dOext(\tsamp)$ depend on $\Nk$ in principle, 
however, for $\Nk \gg 10$ the $\Nk$-dependence is only relevant
for {\em ergodic} systems for which $\sOext \propto 1/\sqrt{\Nk}$ \cite{spmP2} 
and not for the strictly non-ergodic systems we focus on in the present work.
For sampling times $\tsamp$ 
much larger than the typical relaxation time $\taubasin$ of the basins
we have quite generally
\begin{equation}
\left. \begin{array}{ll}
\sOint(\tsamp) \simeq & \sqrt{\taubasin/\tsamp} \\
\sOext(\tsamp) \simeq & \Snonerg
\end{array} \ \right\} \mbox{ for } \tsamp \gg \taubasin
\label{eq_large_tsamp}
\end{equation}
with the ``non-ergodicity parameter" $\Snonerg$ defined by 
\begin{equation}
\Snonerg \equiv \lim_{\tsamp \to \infty} \sOext(\tsamp) \equiv \sOext.
\label{eq_Snonerg_def}
\end{equation}
Note that $\Snonerg > 0$ only holds for strictly non-ergodic systems 
while $\Snonerg=0$ for finite $\taualph$ \cite{spmP2}.
The first asymptotic law in Eq.~(\ref{eq_large_tsamp}) is due to the
$\tsamp/\taubasin$ uncorrelated subintervals
for each $c$-trajectory while the second limit is a consequence of the $\Ocal_c(\tsamp)$ becoming constant.
Equation~(\ref{eq_large_tsamp}) implies
\begin{equation}
\sOtot(\tsamp) \to \Snonerg 
\mbox{ for } \tsamp \gg \Tnonerg \gg \taubasin
\label{eq_sOtot_large_tsamp}
\end{equation}
where the crossover time $\Tnonerg$ to the $\Snonerg$-dominated regime 
is given by $\sOint(\Tnonerg) = \Snonerg$ \cite{spmP2}.
The numerical importance of the inequality $\Tnonerg \gg \taubasin$ is emphasized below.

\subsection{Stationarity}
\label{theo_stationary}

We assume that each $c$-trajectory in its basin is a {\em stationary} stochastic process 
whose joint probability distribution does not change when shifted in time
\cite{VanKampenBook}. 
This may always be achieved by tempering the system over a tempering time $\ttemper \gg \taubasin$.
To take advantage of the stationarity condition
we need to introduce several additional properties.
Let us begin by defining 
the ``empirical sample variance" $v[\xbf]\equiv \Vop^t x_t$ of a time series $\xbf$.
By taking the $k$-average $v_c \equiv \Eop^k v[\xbf_{ck}]$
we obtain the expectation values for each configuration $c$. The expectation value over the
complete $\{ \xbf_{ck} \}$-ensemble is then given by the $c$-average $v \equiv \Eop^c v_c$.
While $\Nc$ and $\Nk$ are assumed to be arbitrarily large, $\Nt$ is in general finite
and for this reason $v_c(\tsamp)$ and $v(\tsamp)$ are {\em apriori} $\tsamp$-dependent
as marked by the arguments. 
The relaxation processes may be characterized using functionals over $\xbf$ 
with a discrete time lag $t$ (with $t=0,\ldots,\Nt-1$) such as the 
``gliding average" \cite{AllenTildesleyBook}
\begin{equation}
c[\xbf;t] = \frac{1}{\Nt-t} \sum_{i=1}^{\Nt-t} x_{i+t} x_i \label{eq_csg}.
\end{equation}
We emphasize that the sum over $i$ is merely done to enhance the statistics
since a stationary stochastic process does not change when shifted in time.
As above we obtain by $k$-averaging the ``autocorrelation function" (ACF)
$c_c(\tau) \equiv \Eop^k c[\xbf_{ck};t]$ 
for a given configuration $c$ and in turn by $c$-averaging
the ACF $c(\tau) \equiv \Eop^c c_c(\tau)$ of the entire $ck$-ensemble. 
It is useful to introduce the differences
\begin{equation}
h_c(\tau) \equiv c_c(0) - c_c(\tau) \mbox{ and } h(\tau) = c(0) - c(\tau). \label{eq_c2h}
\end{equation}
A crucial point is that for stationary processes the
sampling time dependence of $v_c(\tsamp)$ and $v(\tsamp)$
can be traced back to, respectively, $h_c(\tau)$ and $h(\tau)$.
To state this compactly let us introduce the linear operator
\begin{eqnarray}
\Sop[f] & \equiv & \frac{2}{\Nt^2} \sum_{t=1}^{\Nt-t} (\Nt - t) f(t) \label{eq_Sop_def} \\
        & \approx & \frac{2}{\tsamp^2} \int_{0}^{\tsamp-\tau} \ddiff \tau \ (\tsamp - \tau) f(\tau) \label{eq_Sop_cont_def}
\end{eqnarray}
where the first line states the discrete definition and the second line its continuum limit
using that $\tau= t \tincr$ and $\tsamp = \Nt \tincr$.
Note that for $a$ being a constant $\Sop[a]=a$ and this also holds if $f(\tau) \approx a$ 
for a finite but large time window \cite{WKC16,spmP1}.
Following the demonstration given, e.g., in Sec.~2.2 of Ref.~\cite{spmP1}, 
for the ergodic limit it can be seen that the stationarity assumption implies
\begin{equation}
v_c(\tsamp) = \Sop[h_c]
\label{eq_h_c2v_c}
\end{equation}
for each stationary configuration $c$.\footnote{Eq.~(\ref{eq_h_c2v_c}) 
is equivalent to $h_c(\tau) = (\tau^2 v_c(\tau)/2)^{\prime\prime}$
with the prime denoting a derivative with respect to $\tau$.
This relation is closely related to the equivalence of
the Green-Kubo formula and the Einstein relation for transport coefficients \cite{spmP1}.}
Since $[\Eop^c,\Sop]=0$ we have similarly 
\begin{equation}
v(\tsamp) = \Eop^c v_c(\tsamp)= \Sop[\Eop^c h_c] = \Sop[h]
\label{eq_h2v}
\end{equation}
for the $ck$-ensemble. 
The above relations Eq.~(\ref{eq_h_c2v_c}) and Eq.~(\ref{eq_h2v}) imply that $v_c(\tsamp)$ and $v(\tsamp)$ 
must vary strongly for sampling times $\tsamp$ corresponding to strong relaxation processes,
i.e. for times $\tau \approx \tsamp$ where $h_c(\tau)$ and $h(\tau)$ strongly increase.
On the other side $v_c(\tsamp)$ and $v(\tsamp)$ become constant in $\tsamp$-windows
without or with few relaxation processes.
The large time plateau values 
\begin{equation}
h_c \equiv \lim_{\tau \to \infty} h_c(\tau) \mbox{ and } v_c \equiv \lim_{\tsamp \to \infty} v_c(\tsamp)
\label{eq_hpvp}
\end{equation}
(and similarly for $h$ and $v$) are relevant for times exceeding the basin relaxation time $\taubasin$.
It follows from Eq.~(\ref{eq_h_c2v_c}) that $h_c=v_c$ and from Eq.~(\ref{eq_h2v}) that $h=v$.

\begin{figure}[t]
\centerline{\resizebox{.9\columnwidth}{!}{\includegraphics*{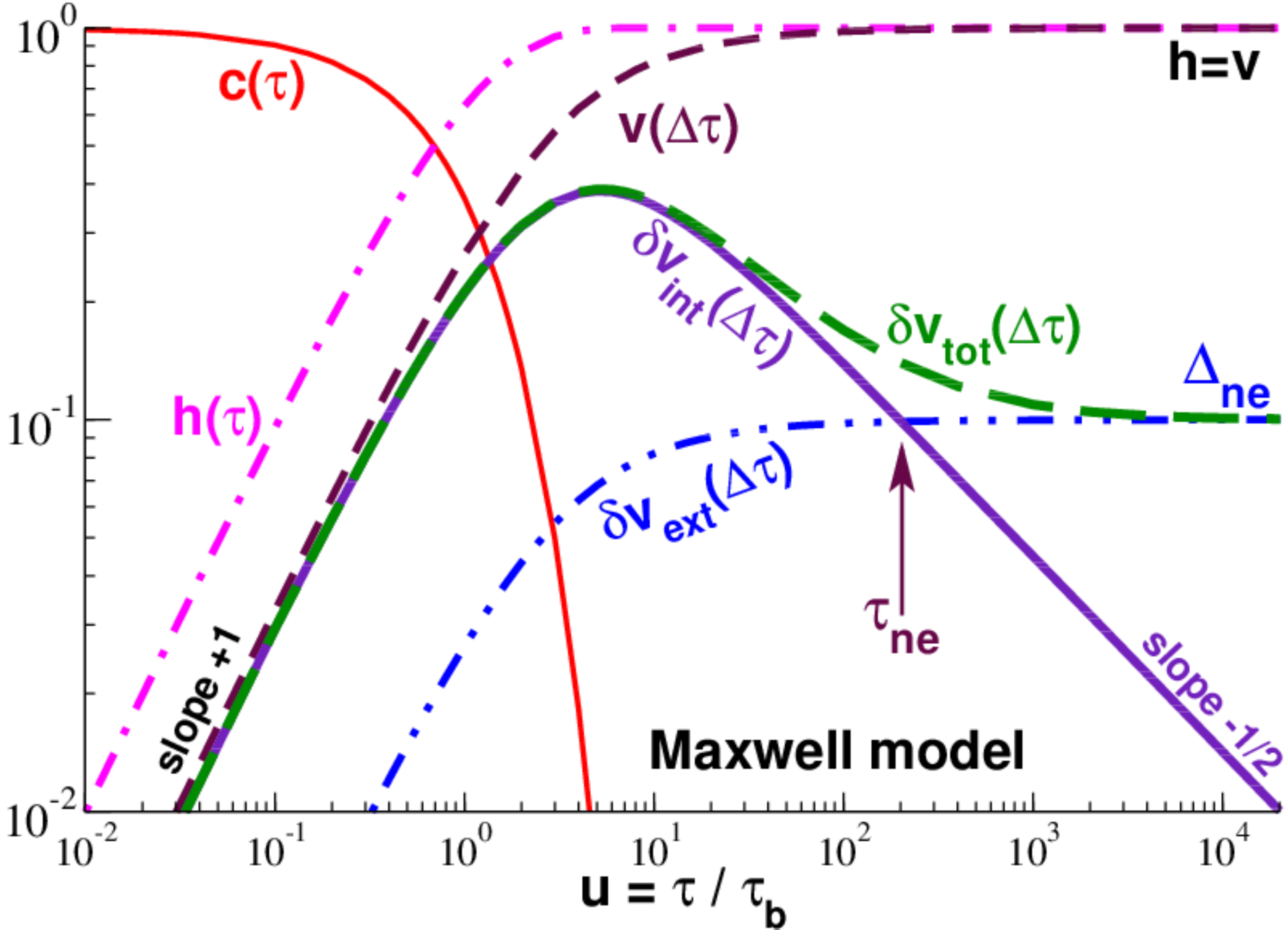}}}
\caption{Illustration of several properties for a Maxwell model.
The ACF is given by $c(\tau)=\exp(-u)$, with $u=\tau/\taubasin$ being the reduced time
and the non-ergodicity parameter $\Snonerg=0.1$. 
We set $h=v=1$ and $\Snonerg/v=0.1$.  $c(\tau)$ is indicated by the thin solid line,
$h(\tau)=c(0)-c(\tau)$ by the thin dot-dashed line,
$v(\tsamp)$ determined according to Eq.~(\ref{eq_v_maxwell}) by the dashed line.
$\svext(\tsamp)$ is obtained using Eq.~(\ref{eq_svext_casestudy}),
$\svint(\tsamp)$ using Eq.~(\ref{eq_dvgauss2dvint_rigor})
and $\svtot(\tsamp)$ using Eq.~(\ref{eq_intro_dOtot}). 
}
\label{fig_maxwell_A}
\end{figure}

We illustrate various points made above by means of a Maxwell (Debye) model 
\cite{spmP1,ChaikinBook,FerryBook},
i.e. we assume a stochastic process with one single exponentially decaying relaxation pathway.
(More generally, response functions and correlation functions of many processes are successfully
fitted by a linear superposition of a finite number or a distribution
of such Maxwell modes \cite{FerryBook}.)
This is presented in Fig.~\ref{fig_maxwell_A}. The ACF $c(\tau)$ of the $ck$-ensemble 
is given by $c(\tau) = \exp(-u)$ as a function of the reduced time $u=\tau/\taubasin$ 
using double logarithmic coordinates.
It follows from Eq.~(\ref{eq_h2v}) that \cite{WKC16,spmP1}
\begin{equation}
v(\tsamp) = 1 - \left[\exp(-\Delta u)-1+\Delta u\right] 2/\Delta u^2
\label{eq_v_maxwell}
\end{equation}
for $\Delta u=\tsamp/\taubasin$.
Let us for simplicity additionally assume
that $h_c(\tau)$ is given by the product of a $c$-dependent constant and
a $c$-independent time-dependence, i.e.
\begin{equation}
h_c(\tau) = p_c h(\tau) \mbox{ with } p_c \ge 0 \mbox{ and } \Eop^c p_c =1.
\label{eq_hc_pc}
\end{equation}
With Eq.~(\ref{eq_h_c2v_c}) and Eq.~(\ref{eq_h2v}) this yields 
$v_c(\tsamp) = p_c v(\tsamp)$. Using Eq.~(\ref{eq_dOext}) we have 
$\dvext(\tsamp) = \left( \Vop^c p_c\right) v(\tsamp)^2$ 
which leads with Eq.~(\ref{eq_Snonerg_def}) to
\begin{equation}
\svext(\tsamp) /\Snonerg = v(\tsamp)/v.
\label{eq_svext_casestudy}
\end{equation}
As seen in Fig.~\ref{fig_maxwell_A} for $\Snonerg=0.1$,
$\svext(\tsamp)$ converges much faster than $\svtot(\tsamp)$ to the common 
large-$\tsamp$ limit $\Snonerg$.

\subsection{Gaussianity}
\label{theo_gaussianity}

As further discussed in Sec.~\ref{sec_V} many non-ergodic stochastic processes
are in fact {\em Gaussian} within each meta-basin.
Using exactly the same arguments put forward in Sec.~3.3 of Ref.~\cite{spmP1} for ergodic
Gaussian stochastic processes it can be shown using Wick's theorem, Eq.~(\ref{eq_app_Wick}),
that $\delta v_c(\tsamp)$ is then given by a functional $\svgauss[h_c]$ of the 
autocorrelation function $h_c(t)$. This functional is defined by \cite{lyuda19a,spmP1}
\begin{eqnarray}
\dvgauss[f] & \equiv & \frac{1}{2\Nt^4} \sum_{i,j,k,l=1}^{\Nt} \ g_{ijkl}^2 \ \mbox{ with } \nonumber \\
g_{ijkl} & \equiv & (f_{i-j} + f_{k-l}) - (f_{i-l} + f_{j-k})
\label{eq_dvgauss_def}
\end{eqnarray}
for any well-behaved function $f(t)$.
Numerically better behaved reformulations of 
Eq.~(\ref{eq_dvgauss_def}) are discussed in Ref.~\cite{spmP1}.
With $a$ and $b$ being real constants we have
\begin{equation}
\svgauss[a]=0 \mbox{ and } \svgauss[b (f-a)] = |b| \ \svgauss[f]
\label{eq_dvgauss_affine}
\end{equation}
and, hence, $\svgauss[h_c]=\svgauss[c_c]$.
Equation~(\ref{eq_dOint}) implies then 
\begin{eqnarray}
\dvint(\tsamp) & =       & \Eop^c \delta v_c^2(\tsamp) = \Eop^c \dvgauss[h_c] \label{eq_dvgauss2dvint} \\
               & \approx & \dvgauss[\Eop^c h_c] =  \dvgauss[h] \label{eq_dvgauss2dvint_approx}
\end{eqnarray}
where the second line is an approximation replacing $h_c(t)$ 
by its $c$-average $h(t)$. This approximation is useful since $h_c(t)$ is not known 
in general, but rather $h(t)$ or $v(\tsamp)$.

Assuming again that Eq.~(\ref{eq_hc_pc}) holds it is seen using the affinity relation 
Eq.~(\ref{eq_dvgauss_affine}) and Eq.~(\ref{eq_svext_casestudy}) that
\begin{equation}
\dvint(\tsamp) = (1 + \epsilon) \ \dvgauss[h] \mbox{ with } \epsilon = \Vop^c p_c.
\label{eq_dvgauss2dvint_rigor}
\end{equation}
Note that commonly $\epsilon \ll 1$, i.e. 
$\svint(\tsamp) \approx \svgauss[h]$ in agreement with Eq.~(\ref{eq_dvgauss2dvint_approx}).
As discussed in Sec.~\ref{sec_V} and Sec.~\ref{sec_spm}, 
$\epsilon \to 0$ for large systems with more or less independent microcells
and the technical assumption Eq.~(\ref{eq_dvgauss2dvint_approx}) thus becomes increasingly rigorous.
Equation~(\ref{eq_dvgauss2dvint_rigor}) is also indicated in Fig.~\ref{fig_maxwell_A} (bold solid line).
We take advantage of the fact that Eq.~(\ref{eq_dvgauss_def}) 
can be solved analytically for the Maxwell model \cite{spmP1}.
An important point is here that $\svint(\tsamp)$ may quite generally become large,
in fact of order of the expectation value $v(\tsamp)$, 
if $\tsamp$ corresponds to a relaxation time of the system. 
This is seen in Fig.~\ref{fig_maxwell_A} by the strong peak of $\svint(\tsamp)$ at $u=\tsamp/\taubasin \approx 6$.
Note also that the total standard deviation $\svtot(\tsamp)$ obtained 
from $\svint(\tsamp)$ and $\svext(\tsamp)$ is given by $\svtot(\tsamp) \approx \svint(\tsamp)$ for $\tsamp \ll \Tnonerg$
and by $\svtot(\tsamp) \approx \svext(\tsamp) \approx \Snonerg$ in the large-$\tsamp$ limit.

\section{System size effects}
\label{sec_V}

\subsection{Phenomenological exponents}
\label{V_exponents}

Stochastic processes of many systems are to a good approximation Gaussian 
since $x_t = \Eop^{\rvec} x_{\rvec t}$ averages over many ($\Nr \gg 1$)
microscopic contributions $x_{\rvec t}$ and the central limit theorem applies \cite{VanKampenBook}. 
Albeit the $x_{\rvec t}$ may be spatially correlated (as discussed below)
the fluctuations commonly decrease with $\Nr$.
As a consequence, $h(\tau)$ and the related variances generally decrease with the system size.
Assuming scale-free correlations one may write \cite{spmP1}
\begin{eqnarray}
h(\tau) \propto v(\tsamp) \propto \svgauss[h] \propto \svint(\tsamp) & \propto& 1/\Nr^{\gaminthat} \label{eq_gamint_def} \\
\svext(\tsamp) \propto \Snonerg & \propto & 1/\Nr^{\gamexthat} 
\label{eq_gamma_def}
\end{eqnarray}
with $\gaminthat$ and $\gamexthat$ being phenomenological exponents. 
That the asymptotic system-size effects for $h(\tau)$ and $v(\tsamp)$ are the same is due to Eq.~(\ref{eq_h2v}).
For Gaussian stochastic processes Eq.~(\ref{eq_dvgauss2dvint_approx})
implies the same exponent $\gaminthat$ for $\svgauss[h]$ and $\svint(\tsamp)$.
As recalled in Ref.~\cite{spmP2} $\gaminthat = 1$ and $\gamexthat = 3/2$
for strictly uncorrelated variables $x_{\rvec}$. 
The uncorrelated reference with $\gaminthat=1$ is often included into 
the definition of the data entries by rescaling 
$x_t$ by a factor proportional to $\sqrt{\Nr}$.\footnote{This
rescaling is not only useful for strictly uncorrelated variables but also for
general fluctuating thermodynamic fields as further discussed in Sec.~\ref{V_T}.}
Hence, $\gaminthat \Rightarrow \gamint \equiv \gaminthat-1$ and 
$\gamexthat \to \gamext \equiv \gamexthat-1$ in the above relations,
i.e.  $\gamint=0$ and $\gamext=1/2$ for {\em rescaled} uncorrelated variables $x_{t\rvec}$.

\begin{figure}[t]
\centerline{\resizebox{.9\columnwidth}{!}{\includegraphics*{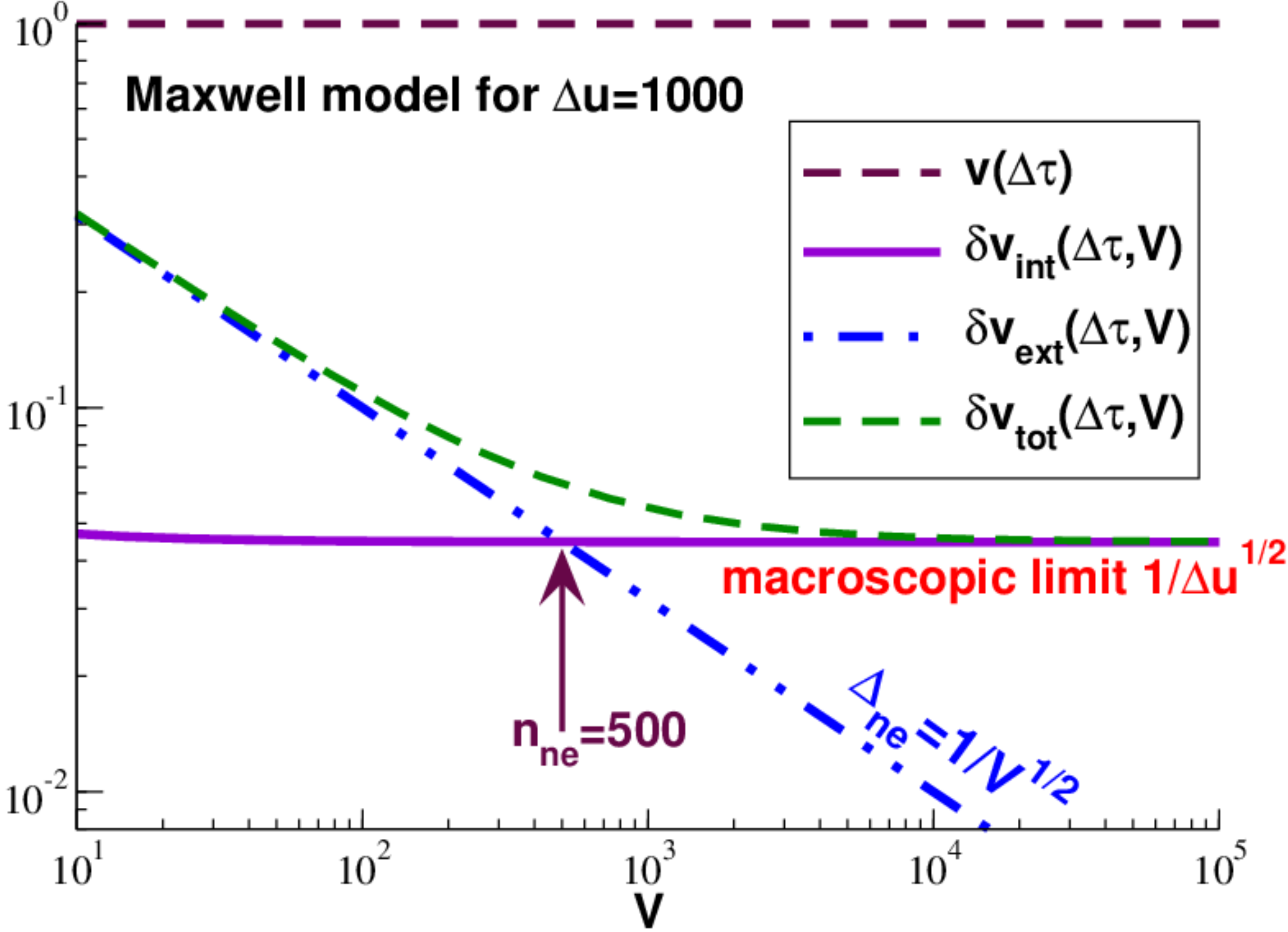}}}
\caption{System-size dependence of $v(\tsamp)$ and the corresponding standard deviations
for the Maxwell model already presented in Fig.~\ref{fig_maxwell_B}. 
It is supposed that $\Snonerg=1/\sqrt{V}$ and $\Delta u = \tsamp/\taubasin=1000$.
Even for such a huge sampling time it would be impossible to fit the correct
exponent $\gamext=1/2$ from the total standard deviation $\svtot(\tsamp,V)$.
}
\label{fig_maxwell_B}
\end{figure}
 
As an example we present in Fig.~\ref{fig_maxwell_B} the system-size dependence of the Maxwell model 
already discussed. We set $\Snonerg=1/\sqrt{V}$ and $V=\Nr$.
Eq.~(\ref{eq_hc_pc}) is again assumed 
and thus in turn also Eq.~(\ref{eq_svext_casestudy}) and Eq.~(\ref{eq_dvgauss2dvint_rigor}).
We focus on one huge reduced sampling time $\Delta u = \tsamp/\taubasin=1000$ 
where $v(\tsamp) \approx v = 1$ and $\svgauss[h] \approx \sqrt{2/\Delta u}$ \cite{spmP1}.
$\svtot(\tsamp,V)$ approaches the macroscopic limit $\svint(\tsamp) \propto V^0$ 
for $V \gg \Nnonerg \approx \Delta u/2$.
Importantly, a crossover regime over at least two orders of magnitude is visible between both
asymptotic limits. This implies that even if $\svtot(\tsamp,V)$ is sampled with a huge constant 
$\tsamp$ an apparent exponent $\gamextapp < \gamext=1/2$ may be measured due to the finite $\tsamp$.
Exponents solely obtained 
from $\svtot(\tsamp=const,V)$ \cite{Procaccia16} may thus be misleading and should be considered with caution.

\subsection{Intensive thermodynamic fields}
\label{V_T}

We now assume that each $c$-trajectory is not only stationary and Gaussian but also at thermal equilibrium
albeit under the constraints imposed to the meta-basin.
We focus on instantaneous {\em intensive} thermodynamic variables $\sigma$ 
(other than the temperature $T$) which are $d$-dimensional volume averages
\begin{equation}
\sigma = \Eop^{\rvec} \sigma_{\rvec} \approx \frac{1}{V} \int \ddiff \rvec \ \sigma_{\rvec}
\label{eq_sigma_Vint}
\end{equation}
over fields $\sigma_{\rvec}$ of same dimension
and $\Nr = V/\dVcell$ being the number of microcells of volume $\dVcell$.
Following the rescaling convention made in Sec.~\ref{V_exponents} 
we use the rescaled variable $x \equiv \sqrt{V} \sigma$.
As already stressed in Ref.~\cite{spmP2}, $\gamint=0$ must even hold for systems with 
long-range correlations if standard thermostatistics can be applied for each basin.
To see this let us remind the reader that the large-$\tsamp$ limit $v_c$ of $v_c(\tsamp)$
is equivalent to the thermodynamically averaged variance of $x$ for the basin.
Using the standard fluctuation-dissipation relation for the fluctuation of intensive thermodynamic variables 
\cite{WXP13,PlischkeBook,Lebowitz67} 
it is then seen that $v_c$ corresponds to a thermodynamic modulus of the $c$-basin
which must be an intensive property, 
i.e. $\gamint=0$.\footnote{For the shear-stress fluctuations considered in Sec.~\ref{sec_simu},
$v_c$ corresponds to the difference $\muFc=\muAc-\mu_c$ of the affine shear modulus $\muAc$ and
the quasi-static shear modulus $\mu_c$ \cite{WXP13,WKC16,lyuda19a,spmP1,spmP2} 
of the configuration $c$ for $\tsamp \to \infty$ with both $\muAc$ and $\mu_c$ being
intensive properties.}
Importantly, the same reasoning {\em cannot} be made for $\gamext$, i.e.  
while $\gamint=0$ must hold $\gamext=1/2$ may not for systems with long-range spatial correlations.
The remainder of the paper illustrates this issue. 

\subsection{Spatial correlations for $\taubasin \ll \tsamp \ll \taualph$}
\label{V_correl}

We have defined above the (generally $\tsamp$-depending) variance of a configuration $c$ by 
$v_c(\tsamp) = \Eop^k v[\xbf_{ck}]$ with $v[\xbf]=\Vop^t x_t$ being the $t$-averaged empirical variance of 
a given time series $\xbf$.
We focus now on {\em static} properties obtained by $k$-averaging over asymptotically 
long $c$-trajectories and assuming $\taubasin \ll \tsamp \ll \taualph$.
In this limit not only the $\tsamp$-dependence of $v_c(\tsamp)$ drops out
but due to the {\em ergodicity within each basin} the {\em time} $t$-average 
can be replaced by an {\em ensemble} $k$-average over the $x_{ck}$ of basin $c$.
We thus lump $t$- and $k$-indices together and the operator $\Eop^k$ replaces $\Eop^k \Eop^t$.
$v_c$ is thus compactly redefined as
\begin{equation}
v_c \equiv \Vop^k x_{ck} 
= \Eop^k x_{ck}^2 - (\Eop^k x_{ck})^2
=  V \Eop^k \delta \sigma_{ck}^2
\label{eq_correl_vc}
\end{equation}
where we have used $\delta \sigma_{ck} = \sigma_{ck} - \Eop^k\sigma_{ck}$ in the last step.
(The prefactor $V$ stems from the rescaling convention.)
Using $\sigma_{ck} = \Eop^{\rvec} \sigma_{ck\rvec}$ and 
$\delta \sigma_{ck\rvec} \equiv \sigma_{ck\rvec} - \Eop^k\sigma_{ck\rvec}$ we write 
\begin{equation}
v_c = V \Eop^k \left(\Eop^{\rvec} \delta \sigma_{ck\rvec} \right)^2
= V \Eop^{\rvec'} \Eop^{\rvec''} \ \underline{\Eop^k \delta \sigma_{ck\rvec'} \delta \sigma_{ck\rvec''}}.
\label{eq_correl_vc_a}
\end{equation}
We define the pair (two-point) correlation function $C_c(\rvec)$ as the average of 
the underlined term in Eq.~(\ref{eq_correl_vc_a}) over all pairs $\rvec'$ and 
$\rvec''=\rvec'+\rvec$. 
Hence,
\begin{equation}
v_c = V \Eop^{\rvec} C_c(\rvec) \approx \int \ddiff \rvec \ C_c(\rvec)
\label{eq_correl_vcB}
\end{equation}
with the first equation stating the discrete sum over all microcells and 
the second relation the corresponding integral for $\dVcell \to 0$.
Hence, $v = \Eop^c v_c = V \Eop^{\rvec} C(\rvec)$ with 
$C(\rvec) \equiv \Eop^c C_c(\rvec)$.\footnote{$v_c \ge 0$ 
sets a constraint on possible $C_c(\rvec)$.
Since $C_c(\rvec)=C_c(r,\nvec)$ depends on the distance $r=||\rvec||$ {\em and} 
the direction $\nvec = \rvec/r$ 
one may write Eq.~(\ref{eq_correl_vcB}) as an $r$-integral of its isotropic
average $C_c^0(r)$ over all $\nvec$.
Due to the imposed (asymptotic) $V$-independence of $v_c$ for all basins (cf. Sec.~\ref{V_T}) 
$C_c^0(r)$ and, hence, $C^0(r) = \Eop^c C_c^0(r)$ must decay more rapidly than $1/r^d$.}

Similarly, $\Dnonerg = \dvext = \Vop^c v_c$ may be rewritten 
exactly as an integral over the four-point correlation function 
$\Eop^c \left[\delta C_c(\rvec_1-\rvec_2) \delta C_c(\rvec_3-\rvec_4)\right]$
using $\delta C_c(\rvec) = C_c(\rvec) - C(\rvec)$.
Unfortunately, without further approximations or physical assumption
this does not yield a useful expression.
One natural route to make progress is to identify a field allowing to express 
$\Dnonerg$ as in integral over a two-point correlation function.
One possible field $\sr$ is obtained by assuming that 
all isotropic and anisotropic contributions to the correlation function 
$C_c(\rvec)$ of the fluctuating field $\sigma_{\rvec}$ rapidly decay on microscopic scales.
We may thus approximate $v_c$, Eq.~(\ref{eq_correl_vc_a}), by the spatial average 
\begin{equation}
v_c \approx \Eop^{\rvec} \scr^2 \mbox{ with } \scr \equiv 
\left(\delta V \Eop^k \delta \sigma_{ck\rvec}^2 \right)^{1/2} 
\label{eq_correl_sr_def}
\end{equation}
being the (rescaled) quenched standard deviation of $\sigma_{c\rvec}$.\footnote{A
simple example is given by a magnetic spin system on a $d$-dimensional lattice subject to a 
strong external quenched magnetic field $H_{\rvec}$ and a weak, say Ising- or Heisenberg-type, 
coupling between neighboring spins \cite{LandauBinderBook,FractalConcepts}.}
(The microscopic field may be renormalized for correlations of finite range.)
As a consequence, $v= \Eop^c v_c \approx \Eop^c \Eop^{\rvec} \scr^2$ and 
$\Snonerg^2 = \Vop^c v_c \approx \Vop^c \Eop^{\rvec} \scr^2$.
Importantly, while the fluctuating field $\sigma_{c\rvec}$ is assumed to be short-ranged,
this does not necessarily imply the same for the $k$-averaged field $\scr$.
An alternative quenched field is discussed in Appendix~\ref{app_X2}.

\section{Simple models}
\label{sec_spm}

\subsection{Introduction}
\label{spm_intro}

We model for analytical and numerical simplicity 
the standard deviations $\scr$ by spatially correlated Gaussian fields $\fcr$
(cf. Appendix~\ref{app_gaussfield} for details), i.e. $\scr = |\fcr|$,
and we focus on (static) moments and correlation functions of these fields.
The approximation Eq.~(\ref{eq_correl_sr_def}) is raised to a postulate, 
i.e. we {\em assume} that $v_c = \Eop^{\rvec} \fcr^2$ and, hence,
\begin{equation}
v   = \mu_2 \equiv \Eop^c\Eop^{\rvec} \fcr^2 \mbox{ and }
\Dnonerg = \Delta_2^2 \equiv \Vop^c \Eop^{\rvec}\fcr^2
\label{eq_spm_postulate}
\end{equation}
hold rigorously. 
More generally,
we denote by $\mu_l \equiv \Eop^c \Eop^{\rvec} \fcr^l$
the total average of the $l$th moment and by $\Delta_l^2 \equiv \Vop^c \Eop^{\rvec} \fcr^l$ 
the corresponding variance. 
Using $\delta \fcr^l \equiv \fcr^l - \mu_l$ we get 
$\Delta_l^2 = \Eop^{\rvec'} \Eop^{\rvec''} \underline{\Eop^c \delta g_{c\rvec'}^l \delta g_{c\rvec''}^l}$. 
With $C_l(\rvec)$ being the average of the underlined term over all pairs $\rvec'$ and 
$\rvec''=\rvec'+\rvec$ this implies
$C_l(\rvec=0) = c_l = \mu_{2l} - \mu_l^2$ and
\begin{equation}
\Delta_l^2 = \Eop^{\rvec} C_l(\rvec) = \frac{1}{V} \int \ddiff \rvec \ C_l(\rvec).
\label{eq_Cl2Deltal}
\end{equation}
Hence, $\Delta_l^2 = c_l \delta V/V$ for spatially uncorrelated fields, 
i.e. for $C_l(\rvec \ne \bfzero) = 0$. 
Importantly, for Gaussian fields $C_{l>1}(\rvec)$ can be expressed 
in terms of $C_1(\rvec)$ and the moments $\mu_l$. Specifically, 
as shown in Appendix~\ref{app_gaussfield},
\begin{equation}
C_2(\rvec) = 2 C_1(\rvec)^2 + 4 \mu_1^2 C_1(\rvec).
\label{eq_Wick3}
\end{equation}
Using Eq.~(\ref{eq_Wick3}) the $\rvec$-average $\Delta_2^2$ over $C_2(\rvec)$, Eq.~(\ref{eq_Cl2Deltal}),
is thus set by $C_1(\rvec)$ and $\mu_1$.
For all model variants discussed below
$C_1(\bfzero) = c_1 = \mu_2 - \mu_1^2$ holds,
i.e. we need to specify additionally either $c_1$ or the moments $\mu_1$ and $\mu_2$.

\begin{figure}[t]
\centerline{
\resizebox{.9\columnwidth}{!}{\includegraphics*{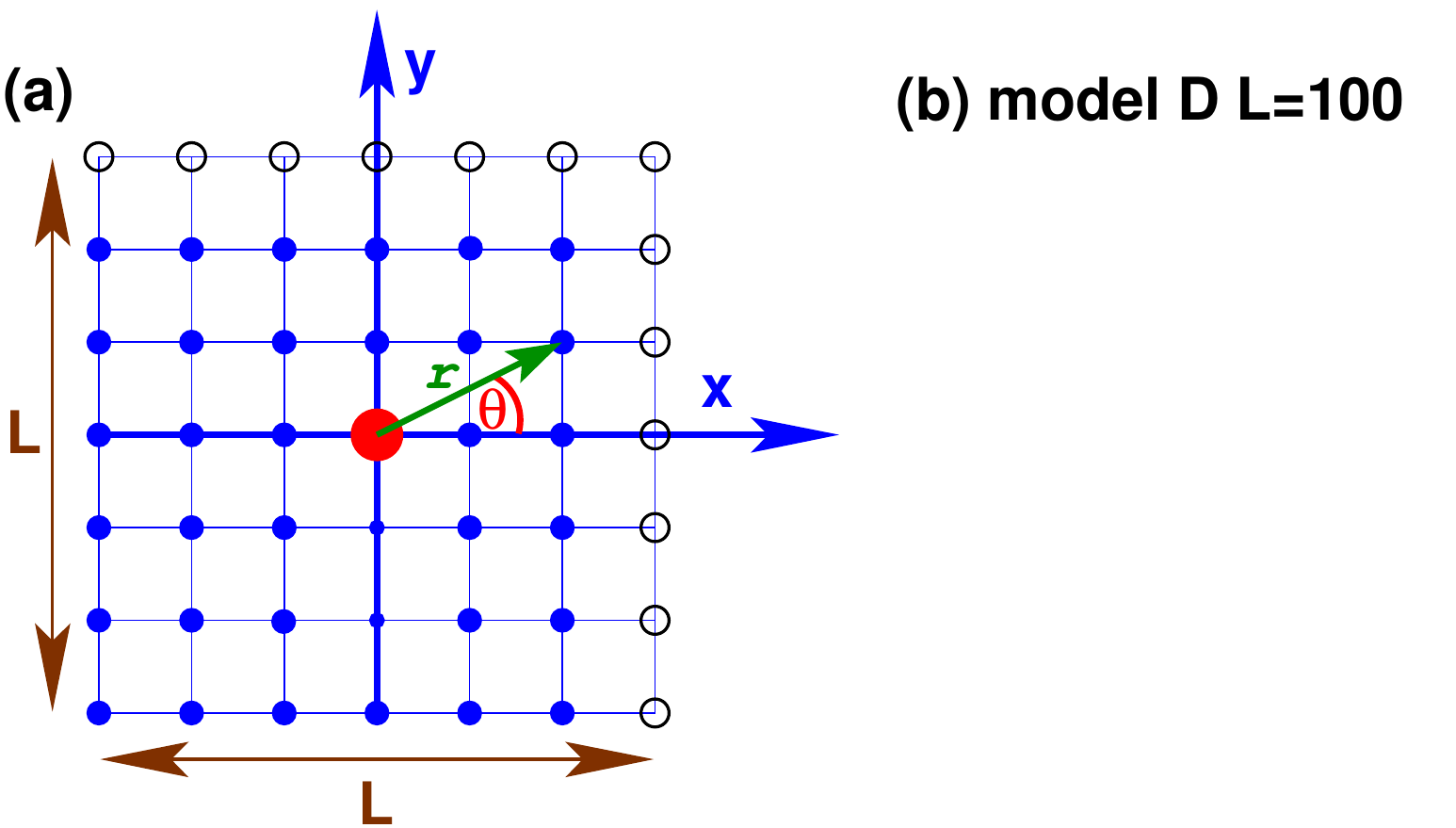}}
}
\vspace*{-4.6cm}\hspace*{4.8cm}
\resizebox{.48\columnwidth}{!}{\includegraphics*{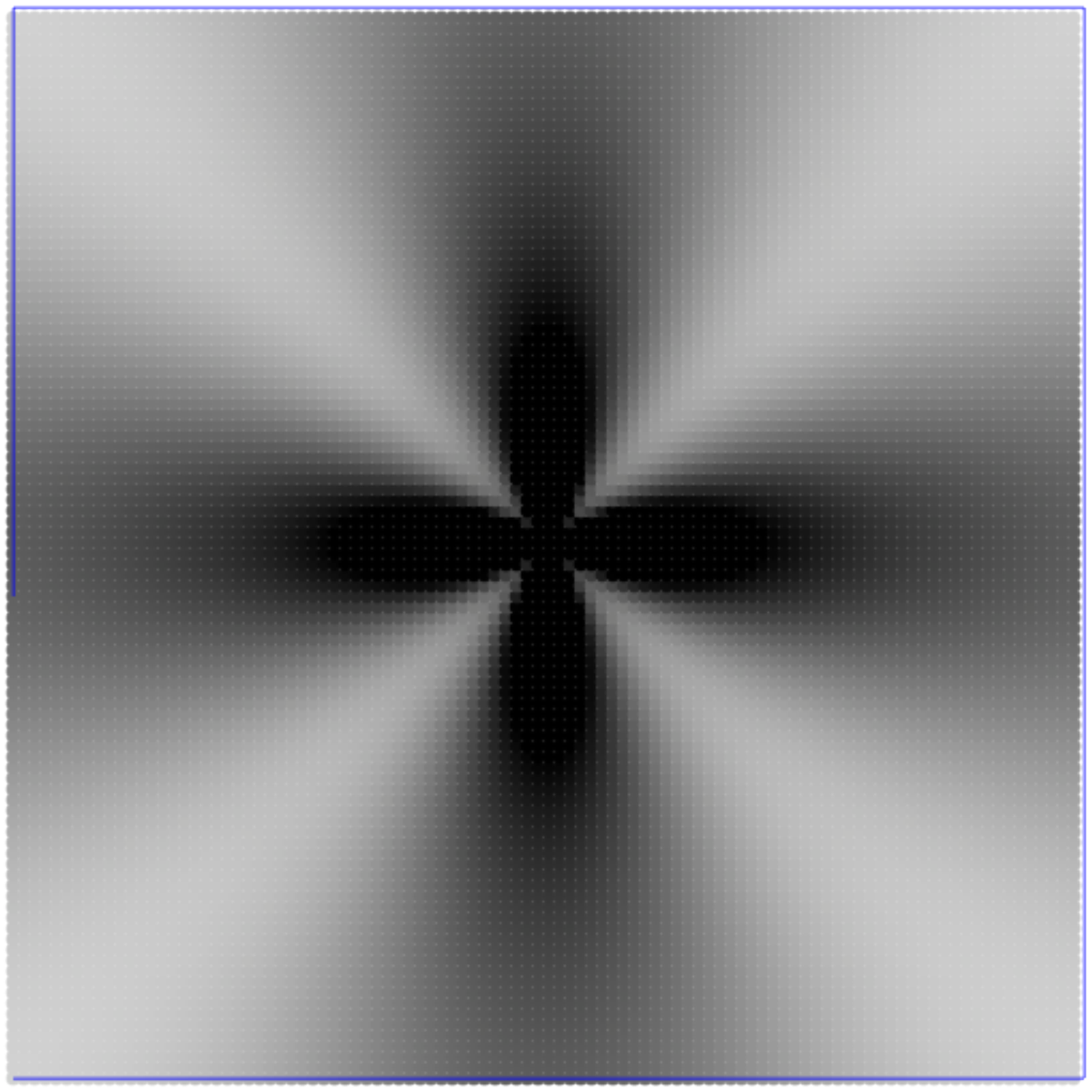}}
\vspace*{-1.5cm}
\caption{Two-dimensional models:
{\bf (a)} sketch of periodic lattice for $L=6$ with filled circles indicating the cells of the principal box
and open circles some periodic images.
For microcells $\rvec$ of the principal box $C_l(\rvec)$ is given by the distance $r$
and the angle $\theta$.
{\bf (b)} $C_1(\rvec)$ for model D with $L=100$, $\alpha_1=0.5$, $\xi=\mu_2=1$, $\mu_1=0$, $\daD=0.5$.
}
\label{fig_Cr_sketch}
\end{figure}

\subsection{Model variants}
\label{spm_models}

As sketched in panel (a) of Fig.~\ref{fig_Cr_sketch} for the two-dimensional case,
we use $d$-dimensional simple cubic lattices of unit lattice 
constant and linear dimension $L$ in all spatial directions, i.e. $\Nr=V=L^d$.
Each of the $\Nr$ lattice sites corresponds to one microcell. 
As usual we use periodic boundary conditions \cite{AllenTildesleyBook}, 
i.e. $\fr$ and the associated correlation functions $C_l(\rvec)$ are $L$-periodic in all spatial directions.
As indicated by filled circles in panel (a) of Fig.~\ref{fig_Cr_sketch},
we focus on the sites of the ``principal simulation box" \cite{AllenTildesleyBook}
characterized by the distance $r=||\rvec||$
from the origin (large filled circle) and the direction $\nvec = \rvec/r$.

We shall consider four model variants.
``Model A" simply assumes that all microcells are uncorrelated, i.e. $C_1(r=0) = c_1$ and $C_1(r>0) =0$.
``Model B" assumes that the correlations decay exponentially
\begin{equation}
C_1(\rvec) = C_1(r) = c_1 \exp(-r/\xi)
\label{eq_modelB}
\end{equation}
with $\xi$ being the correlation length.
Long-range correlations may appear in ``model C" where  
\begin{equation}
C_1(\rvec) = C_1(r) = c_1 (1 + (r/\xi)^2)^{-\alpha_1/2} 
\label{eq_modelC}
\end{equation}
with $\xi$ being again a constant characterizing local physics and $\alpha_1 > 0$.
The shifted power law is used to avoid a divergence at $r=0$ \cite{FractalConcepts}.
Note that $C_1(r) \propto 1/r^{\alpha_1}$ for $r \gg \xi$ and $r \gg 1$.  

Up to now we have assumed that $C_1(\rvec)$ only depends on the
distance $r$ and not on the direction $\nvec$. Interestingly, even for
isotropic systems $C_l(\rvec)$ may depend on $\nvec$ if the stochastic
variable $x(\tau)$ is only a component of a tensor and {\em not} a tensorial invariant.
This is of relevance, e.g., for the shear-stress contribution of the stress tensor 
\cite{Lemaitre14,Fuchs17,lyuda18}.
Focusing on two-dimensional systems and using the angle $\theta$ shown in panel (a) of Fig.~\ref{fig_Cr_sketch}
our ``model D" assumes $C_1(\rvec=\bfzero)=c_1$ and
\begin{equation}
C_1(\rvec) = \tilde{C}_1(r) \left[1  - \daD \cos(4 \theta) \right] \mbox{ for } r > 0
\label{eq_modelD}
\end{equation}
with $\tilde{C}_1(r)$ given by the power-law correlation of model C, Eq.~(\ref{eq_modelC}).
The period $\pi/2$ is due to the fact that $C_l(x,y)$ is even,
i.e. $C_l(x,y)=C_l(-x,y)=C_l(x,-y)=C_l(-x,-y)$, and the assumed equivalence of all spatial directions,
i.e. $C_l(x,y)=C_l(y,x)$.
This is known to hold especially for stress correlations in two-dimensional 
isotropic glasses \cite{Lemaitre14,Fuchs17,lyuda18}.
Interestingly, due to the discrete square lattice the ``anisotropic" term in Eq.~(\ref{eq_modelD}) may give 
finite contributions to the isotropic averages $C_l^0(r)$ over all possible $\theta$ for a given $r$ and (in turn to) 
the sums $\Delta_l^2$, Eq.~(\ref{eq_Cl2Deltal}), over all microcells. There are two reasons for this.
For small $r$ the discrete lattice matters as may be seen by considering the cases $r=1$ or $r=2$.
This effect becomes irrelevant for large $r \gg 1$ and $L$.
More importantly, even for asymptotically large $L$ it matters for small exponents $\alpha_1$
that for $r \ge L/2$ we only sample over microstates in the four corners of the lattice 
around the bisection lines $y = \pm x$ and a correspondingly reduced range of $\theta$ values.
Since $\Delta_l^2 \ge 0$, $\daD$ may not be too negative (depending on the other parameters).
We focus on $\daD = 0.5$.
$C_1(\rvec)$ for $\alpha_1=0.5$, $\mu_2=\xi=1$ and $\mu_1=0$ 
is presented in panel (b) of Fig.~\ref{fig_Cr_sketch}.

\subsection{$\Delta_2$ for Gaussian fields}
\label{spm_Delta2}

We present now $\Delta_2$ for the different models obtained
equivalently by either numerically evaluating Eq.~(\ref{eq_Wick3}) or by explicitly 
first generating random fields (Appendix~\ref{app_gaussfield}) 
and averaging over $\Nc=10^4$ independent configurations.
We focus first on the limit with $\mu_1=0$, i.e. $C_2(\rvec) = 2 C_1(\rvec)^2$,
and set $\mu_2=1$, i.e. $c_1=1$ and $c_2=2$.

\begin{figure}[t]
\centerline{\resizebox{.9\columnwidth}{!}{\includegraphics*{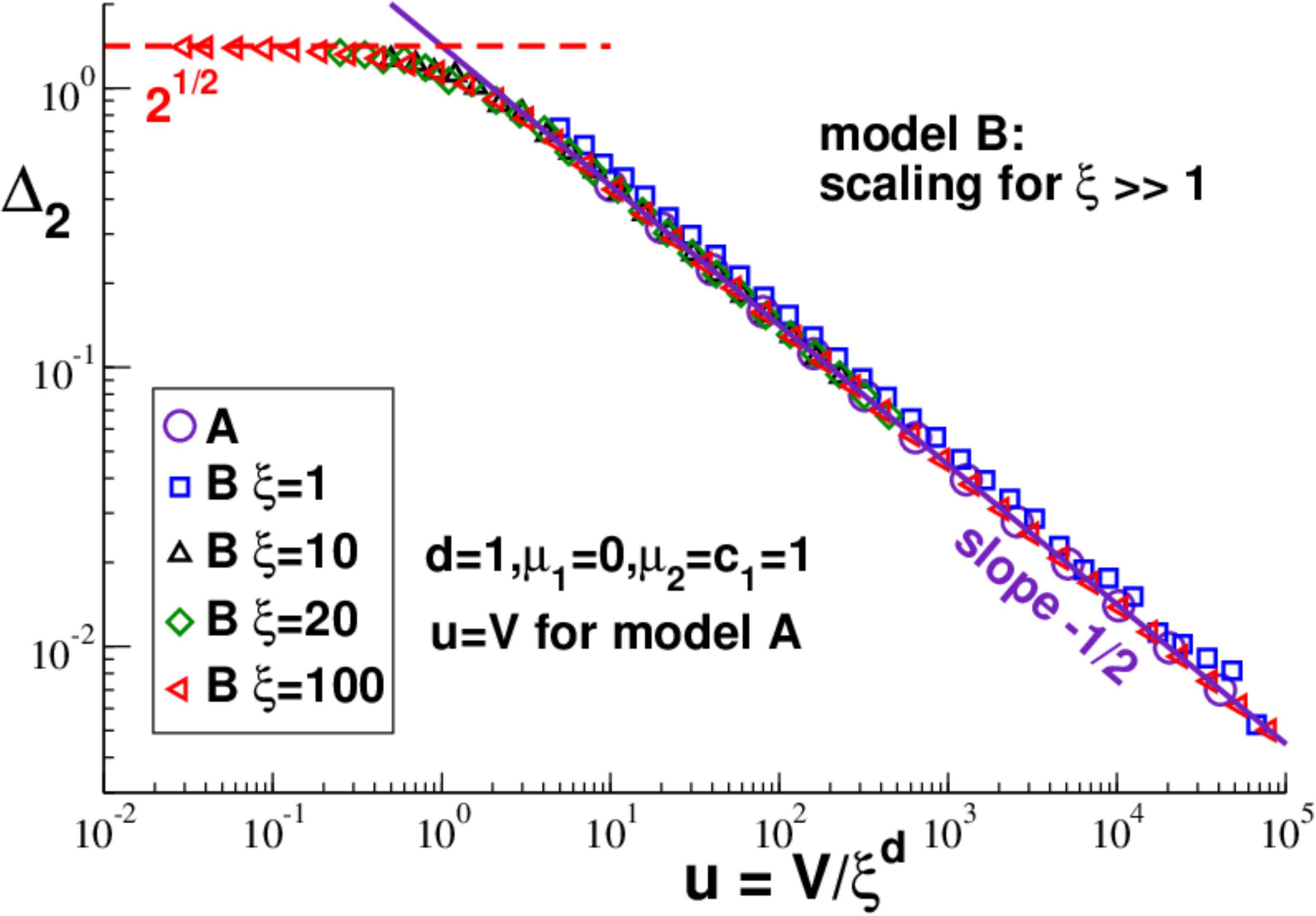}}}
\caption{$\Delta_2$ as a function of the (reduced) system size $u=V/\xi^d$ for models A and B
for $d=1$, $\mu_1=0$ and $\mu_2=1$.
While $\Delta_2 \to \sqrt{2}$ for $u \ll 1$ (dashed horizontal line),
$\Delta_2 \simeq \sqrt{2/u}$ for sufficiently large systems (bold solid line).
}
\label{fig_Delta2_AB}
\end{figure}
If $C_1(\rvec)$ and thus $C_2(\rvec)$ are short-ranged, the system-size must
become rapidly irrelevant and, hence, 
\begin{equation}
\Delta_2 \simeq 1/V^{\gamext} \mbox{ with } \gamext = 1/2.
\label{eq_Delta2_gamext_uncorr}
\end{equation}
This behavior is shown in Fig.~\ref{fig_Delta2_AB}
for different one-di\-men\-sion\-al ($d=1$) systems.
Since for model A $C_2(\bfzero)=2$ and $C_2(\rvec \ne \bfzero)=0$,
this implies $\Delta_2 = \sqrt{2/V}$ as indicated by the bold solid line.
As one expects the data for model B scales if traced as a function of the reduced volume $u=V/\xi^d$.
Naturally, the scaling is not perfect for small $\xi$ due to the discrete lattice.
For $u \ll 1$ we have $C_2(\rvec) \approx 2$ according to Eq.~(\ref{eq_Wick3}).
As shown by the dashed horizontal line we thus have $\Delta_2 \to \sqrt{2}$
for $u \ll 1$ while in the opposite limit $\Delta_2 \simeq \sqrt{2/u}$, as expected.

\begin{figure}[t]
\centerline{\resizebox{.9\columnwidth}{!}{\includegraphics*{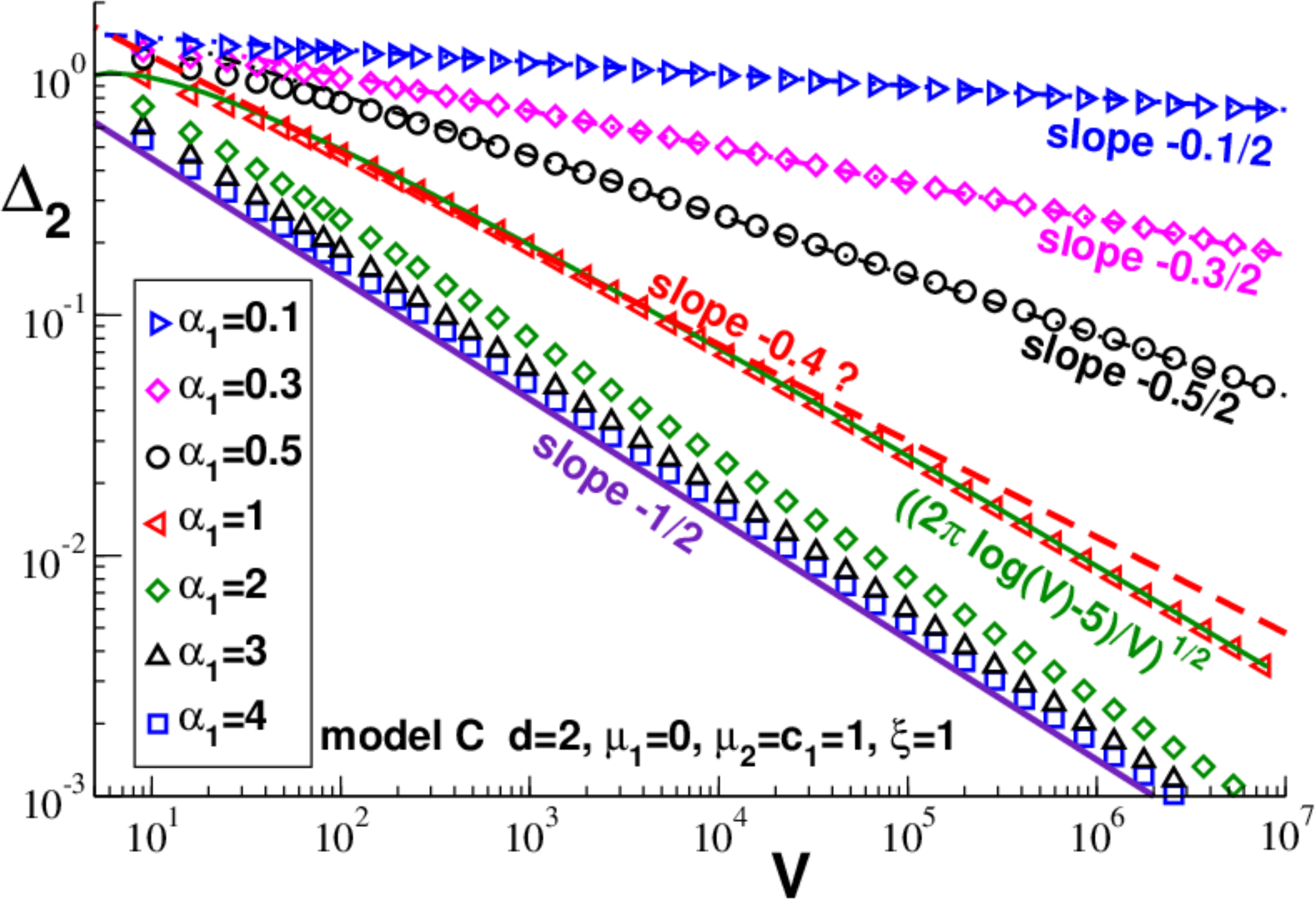}}}
\caption{$\Delta_2(V)$ for model C for $d=2$, $\mu_1=0$, $\mu_2=1$, $\xi=1$ and 
different power-low exponents $\alpha_1$ as indicated.
While $\gamext=1/2$ for $\alpha_2 > d$ (bold solid line) and $\gamext = \alpha_1/d$ for $\alpha_d< d$ (dash-dotted lines),
an apparent exponent $\gamextapp \approx 0.4$ (bold dashed lines) is observed for $\alpha_2=d$.
The thin solid line indicates the logarithmic correction for $\alpha_2 \to d$.
}
\label{fig_Delta2_C}
\end{figure}

Long-range correlations may appear in model C as seen in Fig.~\ref{fig_Delta2_C}.
Since $\mu_1=0$ we have $\alpha_2 = 2 \alpha_1$ in the large-$r$ limit.
Depending on the value of $\alpha_2$ and the spatial dimension $d$
it is readily seen from Eq.~(\ref{eq_Cl2Deltal}) that 
$\gamext = 1/2$ for $\alpha_2 > d$,
while in the opposite limit
\begin{equation}
\gamext = \alpha_2/2d \ \mbox{ for } \alpha_2 < d.
\label{eq_modelC_long_ranged}
\end{equation}
Both relations are seen to hold in Fig.~\ref{fig_Delta2_C} for two-dimensional systems ($d=2$).
The cases with long-range correlations are emphasized by dash-dotted lines. 
For large $\alpha_1$ we see that $\Delta_2$ approaches
the limit $\Delta_2 = \sqrt{2/V}$ (bold solid lines) of uncorrelated microcells (model A).
Since
\begin{equation}
\Delta_2 \simeq (\log(V)/V)^{1/2} \mbox{ for } \alpha_2 = d
\label{eq_modelC_log_ranged}
\end{equation}
we observe strong curvature for $\alpha_1=1$.
Moreover, this limiting case is rather well fitted over at least two orders of magnitude
by an apparent power law (bold dashed lines) with $\gamextapp \approx 0.4$.
(Similar results are obtained in other dimensions.)
This demonstrates (if yet necessary) that such power-law fits should be treated with care.
The thin solid line shows a logarithmic fit suggested by Eq.~(\ref{eq_modelC_log_ranged}).

\begin{figure}[t]
\centerline{\resizebox{.9\columnwidth}{!}{\includegraphics*{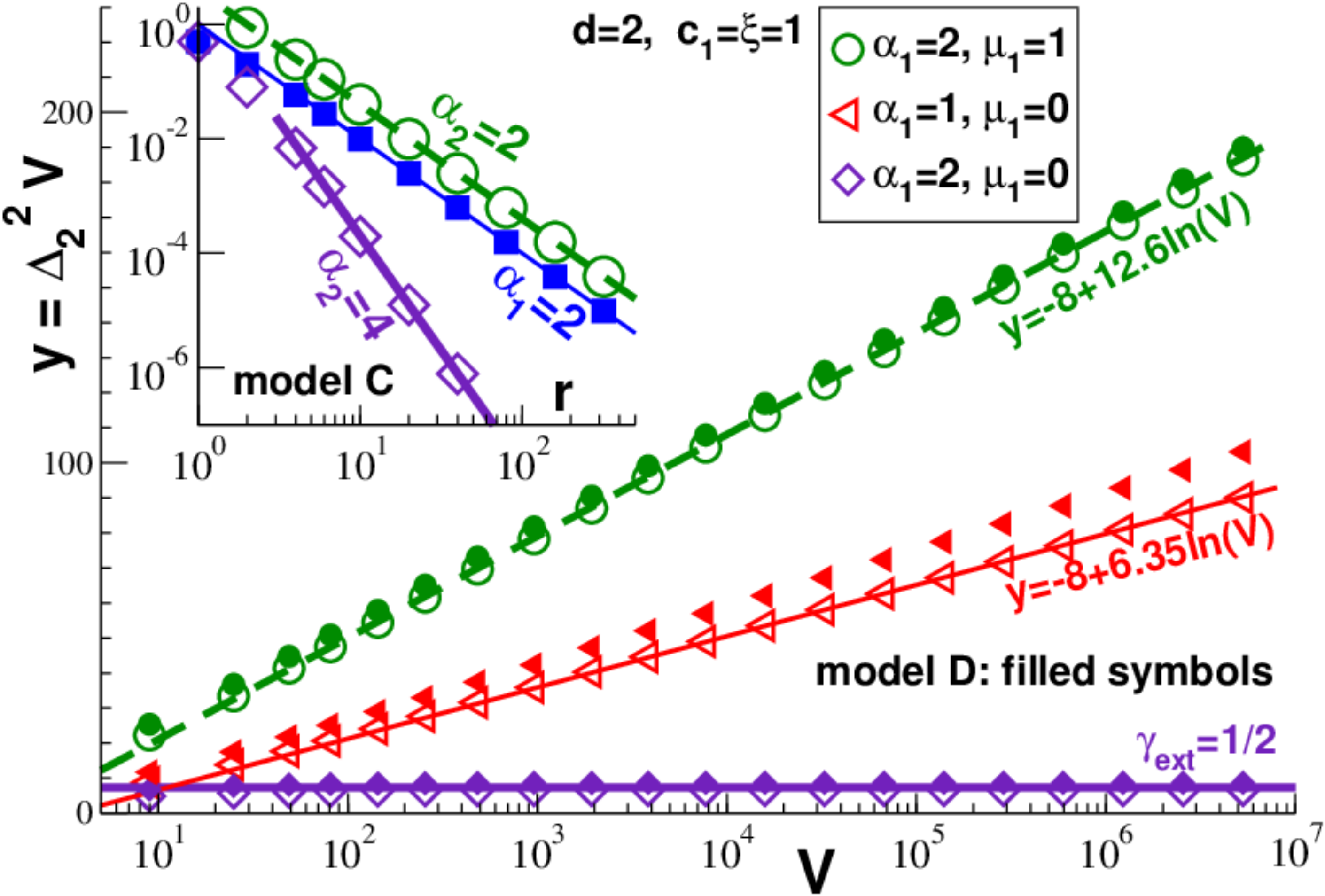}}}
\caption{Main panel: $y=\Delta_2^2V$ {\em vs.} $V$ for model C (open symbols) and model D (filled symbols).
As shown by the bold horizontal line we have $\gamext=1/2$ for $\alpha_1=2$ if $\mu_1=0$
while $y$ increases linearly for both models for $\mu_1=1$.
Inset: Double-logarithmic representation of $C_1(\rvec)$ (filled squares)
and $C_2(\rvec)$ (open symbols) for model C with $\alpha_1=2$ and $\mu_1=0$ (diamonds)
and $\mu_1=1$ (circles). 
We have $C_2(\rvec)=2/r^4$ in the former case (bold solid line) and $C_2(\rvec)=4/r^2$ (dashed line) 
in the latter.
}
\label{fig_Delta2_C2mu1}
\end{figure}

Up to now we have set $\mu_1=0$, i.e. $\alpha_2=2\alpha_1$ for model C and model D.
$\alpha_1=2$ thus implies $\alpha_2=4$, i.e. long-range correlations are irrelevant for $d=2$.
This may be better seen using the half-logarithmic coordinates in the main panel of 
Fig.~\ref{fig_Delta2_C2mu1} where $y =\Delta_2^2 V$ is plotted as a function of $V$.
Indeed the data for $\mu_1=0$ and $\alpha_1=2$ (diamonds) are strictly horizontal (bold solid line)
and logarithmic corrections only appear for $\alpha_1=1$ (triangles).\footnote{The 
differences between models C and D for small $\alpha_2$ and large $L$ are due to the contributions
of the anisotropic term of model D for discrete square lattices.}
Interestingly, the linear-$C_1(\rvec)$-contribution in Eq.~(\ref{eq_Wick3}) is readily switched on
using a finite $\mu_1$ and, as can be seen for model C in the inset of Fig.~\ref{fig_Delta2_C2mu1},
$C_2(\rvec) \approx 4 \mu_1 C_1(\rvec)$ already for small $\mu_1$, i.e. $\alpha_2 \approx \alpha_1$.
As shown in the main panel, $\Delta_2$ reveals 
strong logarithmic behavior (circles).

\section{Mapping on shear-stress data}
\label{sec_simu}

It is tempting to tune the parameters of model C or D to fit the corresponding data obtained
for shear-stresses in simulated model glasses \cite{WXP13,Procaccia16,lyuda19a,spmP1,spmP2}.
We focus on systems formed by polydisperse Lennard-Jones (pLJ) particles in two dimensions.
See Refs.~\cite{WXP13,spmP1} for a description of the Hamiltonian,
the simulation method, the quench protocol and thermodynamic and structural properties.
Boltzmann's constant and the average particle diameter are set to unity
and Lennard-Jones units \cite{AllenTildesleyBook} are used.
We impose a temperature $T=0.2$ --- much smaller than the glass transition temperature 
$\Tglass \approx 0.26$ \cite{WXP13,spmP1} --- and sample $\Nc=100$ independent configurations 
containing between $n=100$ and $n=40000$ particles. 
The number density is essentially system-size independent and of order unity,
i.e. the particle number $n$ and the volume $V$ are numerically similar.
\begin{figure}[t]
\centerline{\resizebox{.9\columnwidth}{!}{\includegraphics*{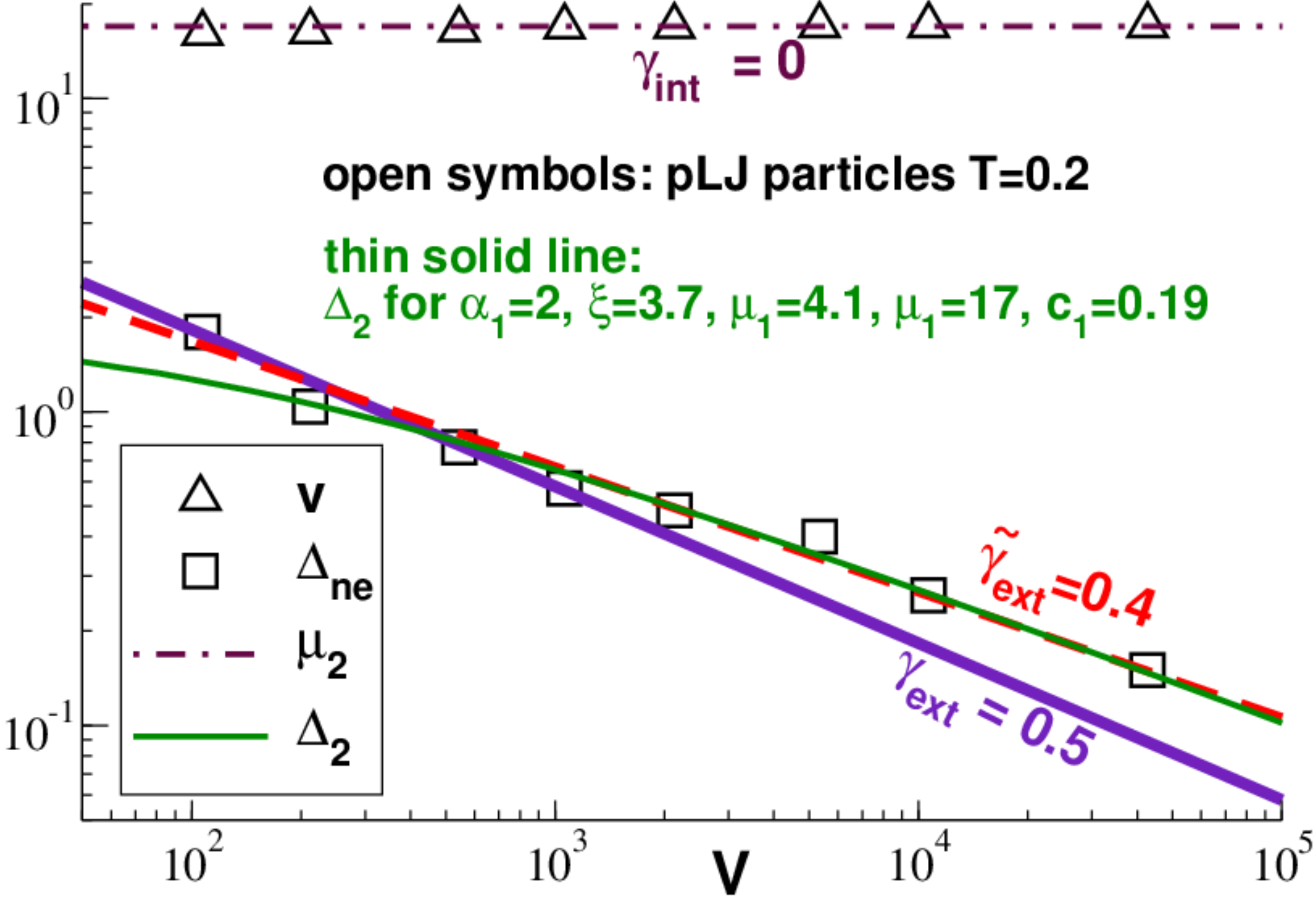}}}
\caption{Mapping of model C on data obtained from the shear-stress fluctuations of MC simulations
of pLJ particles in two dimensions (open symbols).
As can be seen $v \approx 17$ is $V$-independent, i.e. $\gamint=0$.
Imposing $\alpha_1=2$ and $\mu_2=v$ and fitting $\mu_1=4.1$ and $\xi=3.7$ yields $\Delta_2$ (thin solid line).
}
\label{fig_spm2simu}
\end{figure}
The only observable relevant for the present work is the shear-stress contribution $\sigma$
to the (excess) stress tensor \cite{AllenTildesleyBook,spmP1,spmP2}.
Measurements are performed each MC step over a total sampling time $\tsampmax=10^7$.
The stochastic process $x(\tau)$ is obtained as suggested by the convention made in 
Sec.~\ref{V_exponents} by rescaling $\sigma \Rightarrow x_t \equiv \sqrt{V/T} \ \sigma$.
As described in Sec.~\ref{theo_ck}, we obtain from $v[\xbf_{ck}]$ 
the expectation value $v(\tsamp)$, the total variance $\dvtot(\tsamp)$
and its contributions $\dvint(\tsamp)$ and $\dvext(\tsamp)$
and, finally, $\Snonerg$ from the long-time limit of $\svext(\tsamp)$, Eq.~(\ref{eq_Snonerg_def}).
The $\tsamp$- and $\Nk$-dependences are described in Ref.~\cite{spmP2}.

The data for $v(\tsamp = 10^6) \approx v$ and $\Snonerg$ are presented in Fig.~\ref{fig_spm2simu}. 
Since $v$ is the (rescaled) fluctuation of the shear stress, 
it is (essentially) $V$-independent, i.e. $\gamint=0$ in agreement with Sec.~\ref{V_T}.
As already emphasized elsewhere \cite{spmP1,spmP2}, 
$\Snonerg$ does not decay with an exponent $\gamext=1/2$ (bold solid line)
but with a weaker apparent exponent $\gamextapp \approx 0.4$ (bold dashed line).
This finding is qualitatively in agreement with the same exponent observed in Sec.~\ref{sec_spm}
for either $\alpha_1=1$ and $\mu_1=0$ (Fig.~\ref{fig_Delta2_C}) or
$\alpha_1=2$ and $\mu_1=1$ (Fig.~\ref{fig_Delta2_C2mu1}).
Naturally, $\mu_2$ is set by the $V$-independent value of $v \approx 17$
as indicated by the dash-dotted horizontal line.
Motivated by recent theoretical and numerical work \cite{Lemaitre14,Fuchs17,lyuda18} 
we impose $\alpha_1=d =2$ and fit the remaining parameters $\xi$ and $\mu_1$ of model C.
This yields with $\xi \approx 3.7$ and $\mu_1 \approx 4.1$ a nice fit (thin solid line) of $\Snonerg$ for $n > 100$.
Similar values have been found for model D.
The main limitation for a more critical test of the mapping 
is that $\Snonerg$ is not known for larger system sizes.

\section{Conclusion}
\label{sec_conc}

Extending recent work \cite{spmP1,spmP2} the present study focused on the expectation value 
$v(\tsamp)$ and the standard deviations $\svtot(\tsamp)$, $\svint(\tsamp)$ and $\svext(\tsamp)$
of the empirical variance $v[\xbf]$ of time series $\xbf$, Eq.~(\ref{eq_xbf_def}), 
of strictly non-ergodic stochastic processes recorded over a sampling time $\tsamp$.
  
Our first aim was to give an uncluttered summary (Sec.~\ref{sec_theo}) 
of some useful notations (Sec.~\ref{theo_definitions})
and general relations important for the characterization of ensembles
$\{\xbf_{ck}\}$ of such time series.
At variance to ergodic processes the external standard deviation $\svext(\tsamp)$
becomes for non-ergodic systems constant, $\svext(\tsamp) \simeq \Snonerg > 0$, 
for large $\tsamp \gg \taubasin$, and thus in turn so does also the total standard deviation 
$\svtot(\tsamp)$ for $\tsamp \gg \Tnonerg(V) \gg \taubasin$, Eq.~(\ref{eq_sOtot_large_tsamp}).
  
Our second aim was to emphasize by means of a simple analytically
feasible example (Figs.~\ref{fig_maxwell_A} and \ref{fig_maxwell_B})
that it is therefore questionable to numerically determine the 
sy\-stem-size exponent $\gamext$ of $\Snonerg(V)$ uniquely from 
the total standard deviation $\svtot(\tsamp=const,V)$.
We argued that one should rather analyze the more rapidly converging 
$\svext(\tsamp,V)$ both with respect to $\tsamp$ and $V$. 
 
Our third aim was to better understand the system-size dependence 
of the static properties $v$ and $\Snonerg$ for $\tsamp \gg \taubasin$
in systems with correlated microcells (Sec.~\ref{sec_V}).
We have thus investigated in Sec.~\ref{sec_spm} simple models where the 
(unaveraged) fluctuating microscopic contributions $\sigma_{\rvec}$ 
are essentially decorrelated but their (rescaled) $k$-averaged standard deviation $\sr$ may not.
For simplicity these frozen fields $\sr = |\fr|$
were modelled by spatially correlated Gaussian fields $\fr$ 
(Appendix~\ref{app_gaussfield}).
$v$ and $\Snonerg$ are given, respectively, by the moments $\mu_2$ and $\Delta_2$ of the $\fr$-field,
Eq.~(\ref{eq_spm_postulate}).
We have thus expressed $\Snonerg$ in terms of an effective two-point correlation function 
$C_2(\rvec)$, Eq.~(\ref{eq_Cl2Deltal}).
For consistency with general intensive thermodynamic fields (Sec.~\ref{V_T}), 
$\mu_2$ is set to be $V$-independent ($\gamint=0$).
As seen in Fig.~\ref{fig_Delta2_AB} for models A and B and in Fig.~\ref{fig_Delta2_C} for model C 
with $\alpha_2 > d$, $\Delta_2 \propto 1/V^{\gamext}$ with $\gamext =1/2$ only holds for sufficiently 
strongly decreasing spatial correlations. 
Logarithmic corrections become relevant for models C and D for $\alpha_2 \to d$
where $\Delta_2$ may be fitted over two orders of magnitude by an apparent 
exponent $\gamextapp \approx 0.4$ (Fig.~\ref{fig_Delta2_C}).
A similar alternative approach yielding numerically equivalent results 
is mentioned in Appendix~\ref{app_X2}.

Our fourth aim was to point out (Sec.~\ref{sec_simu}) that rather similar
behavior is observed for shear-stress fluctuations in amorphous glasses \cite{spmP2}. 
By insisting on $\alpha_1=d=2 \approx \alpha_2$ and tuning the parameters $\mu_1$ 
and $\xi$ of models C or D it was possible to fit the data (Fig.~\ref{fig_spm2simu}). 
This finding suggests the observed apparent exponent $\gamextapp \approx 0.4$ 
\cite{spmP1,spmP2,Procaccia16} to be due to {\em marginally} long-range correlations of 
quenched shear-stress fluctuations.
  
Obviously, this does not necessarily imply that other aspects of the stress correlations 
in these systems are captured by simple models based on the key postulate 
Eq.~(\ref{eq_spm_postulate}) and, especially,
on the technical relation Eq.~(\ref{eq_Wick3}) assuming correlated Gaussian fields.
To clarify this issue future work \cite{spmP4} will focus on the characterization of the 
spatial correlations of different quenched fields such as the 
``local covariance field" $v_{cr} = V \Eop^k \delta \sigma_{ck\rvec} \delta \sigma_{ck}$ 
(Appendix~\ref{app_X2}) which may be constructed from the shear stress fields $\sigma_{ck\rvec}$
numerically obtained following Lema\^\i tre \cite{Lemaitre14}.
 
For simplicity of the presentation we have assumed in the present work a diverging 
longest system relaxation time $\taualph$ albeit for real physical, biological or socio-eco\-no\-mical 
systems $\taualph$ is generally finite. Importantly, $\svext(\tsamp)$ must vanish in the ergodic limit 
for $\tsamp \gg \taualph$. It is appropriate for systems with a sluggish glass-like dynamics to
redefine $\Snonerg$ as the intermediate plateau value of $\svext(\tsamp)$.
Naturally, all the presented results hold as long as $\tsamp$, $\taubasin$ and $\Tnonerg$
are much smaller than $\taualph$.  The complete description of 
the standard deviations $\svtot(\tsamp,\Nc,\Nk)$, $\svint(\tsamp,\Nc,\Nk)$ and $\svext(\tsamp,\Nc,\Nk)$ 
is more intricate. See Sec.~4.6 of Ref.~\cite{spmP2} for some first results.

\section*{Author contribution statement}
JB, ANS and JPW designed the project.
GG, LK and JPW performed the simulations.
JPW wrote the manu\-script benefitting from contributions of all authors.

\section*{Acknowledgments}
We acknowledge computational
resources from the HPC cluster of the University of Strasbourg.

\appendix

\section{Correlation functions of Gaussian fields}
\label{app_gaussfield}

Let $y_i$ be a normal distributed random field of zero mean, i.e. $\langle y_i \rangle = \mu_1 = 0$,
with $i$ standing for the discrete time or spatial position. 
($\langle \ldots \rangle$ stands here for a $c$-average over $\Nc \to \infty$ independent configurations.)
Wick's theorem \cite{VanKampenBook} thus holds, i.e.
\begin{eqnarray}
\hspace*{-.8cm}\la y_i y_j y_k y_l \ra & = & \nonumber\\
& & \hspace*{-2.cm} 
\la y_i y_j \ra \la y_k y_l \ra 
+ \la y_i y_k \ra \la y_j y_l \ra
+ \la y_i y_l \ra \la y_j y_k \ra.
\label{eq_app_Wick}
\end{eqnarray}
This implies in turn
$\langle y_i^2 y_j^2 \rangle - \langle y_i^2 \rangle \langle y_j^2 \rangle = 2 \langle y_i y_j \rangle^2$.
With the indices corresponding to spatial positions and assuming translational invariance 
this shows that $C_2(\rvec) = 2 C_1(\rvec)^2$ for $\mu_1 = 0$.
If we consider instead the field $g_i = y_i + \mu_1$
with finite first moment 
$\mu_1 = \langle g_i \rangle$, 
$C_1(\rvec)$ remains unchanged while $C_2(\rvec)$ does not.
By substituting $y_i=g_i-\mu_1$, 
expanding $C_2(\rvec)$ and using the invariance of $C_1(\rvec)$ it is seen that more generally
Eq.~(\ref{eq_Wick3}) holds.\footnote{It 
is used here that 
$\langle s_i^2 s_j \rangle - \langle s_i^2 \rangle \langle s_j \rangle =  2 \mu_1 C_1(\rvec)$.}

Random Gaussian fields $\fr$ corresponding to the models of 
Sec.~\ref{spm_models} have been explicitly generated numerically and we have 
measured the correlation functions $C_l(\rvec=\rvec''-\rvec')$ by averaging 
(consistently with the periodic boundary conditions) over all pairs of cells 
$\rvec'$ and $\rvec''$ and the $\Nc$ independent configurations.
Model A is trivially obtained by generating for each configuration
$\Nr$ uncorrelated normal-distributed random numbers $\zeta_{\rvec}$
of zero mean and unit variance and setting
$\fr = \mu_1 + a_0 \zeta_{\rvec}$ with $c_1 = a_0^2$.
Spatially correlated random numbers $y_{\rvec}=\fr-\mu_1$ are obtained by setting
$y_{\rvec'} = \sum_{\rvec''} a_{\rvec'\rvec''} \zeta_{\rvec'}$
where the ``response" matrix $a_{\rvec'\rvec''}$ 
is uniquely determined by the imposed correlation function $C_1(\rvec)$ as shown below.
Importantly, $\fr=y_{\rvec}+\mu_1$ is thus a linear superposition of Gaussian variables 
and therefore also Gaussian.\footnote{This implies that $\fr$ may be negative even
for large $\mu_1 > 0$ while the standard deviation $\sr$ (cf. Sec.~\ref{V_correl})
of the microscopic field $\sigma_{\rvec}$ must be positive definite.}
As a consequence Eq.~(\ref{eq_Wick3}) applies.
Following a standard procedure \cite{FractalConcepts} 
one way to obtain the $y_{\rvec}$-fields is to compute in turn
the Fourier transforms $\zeta_{\qvec} = \Fcal[\zeta_{\rvec}]$ and $C_1(\qvec) = \Fcal[C_1(\rvec)]$,
the product $y_{\qvec} = \sqrt{C_1(\qvec)} \zeta_{\qvec}$,
and finally the inverse Fourier transform $y_{\rvec} = \Fcal^{-1}[y_{\qvec}]$
with $\Fcal$ standing for the $d$-dimensional discrete Fourier transform 
and $\Fcal^{-1}$ for its inverse. It is used here that $\zeta_{\qvec} \zeta_{-\qvec} = 1$
and that $C_1(\qvec)$ is real, even, positive and commensurate with the simulation box.
\begin{figure}[t]
\centerline{\resizebox{.9\columnwidth}{!}{\includegraphics*{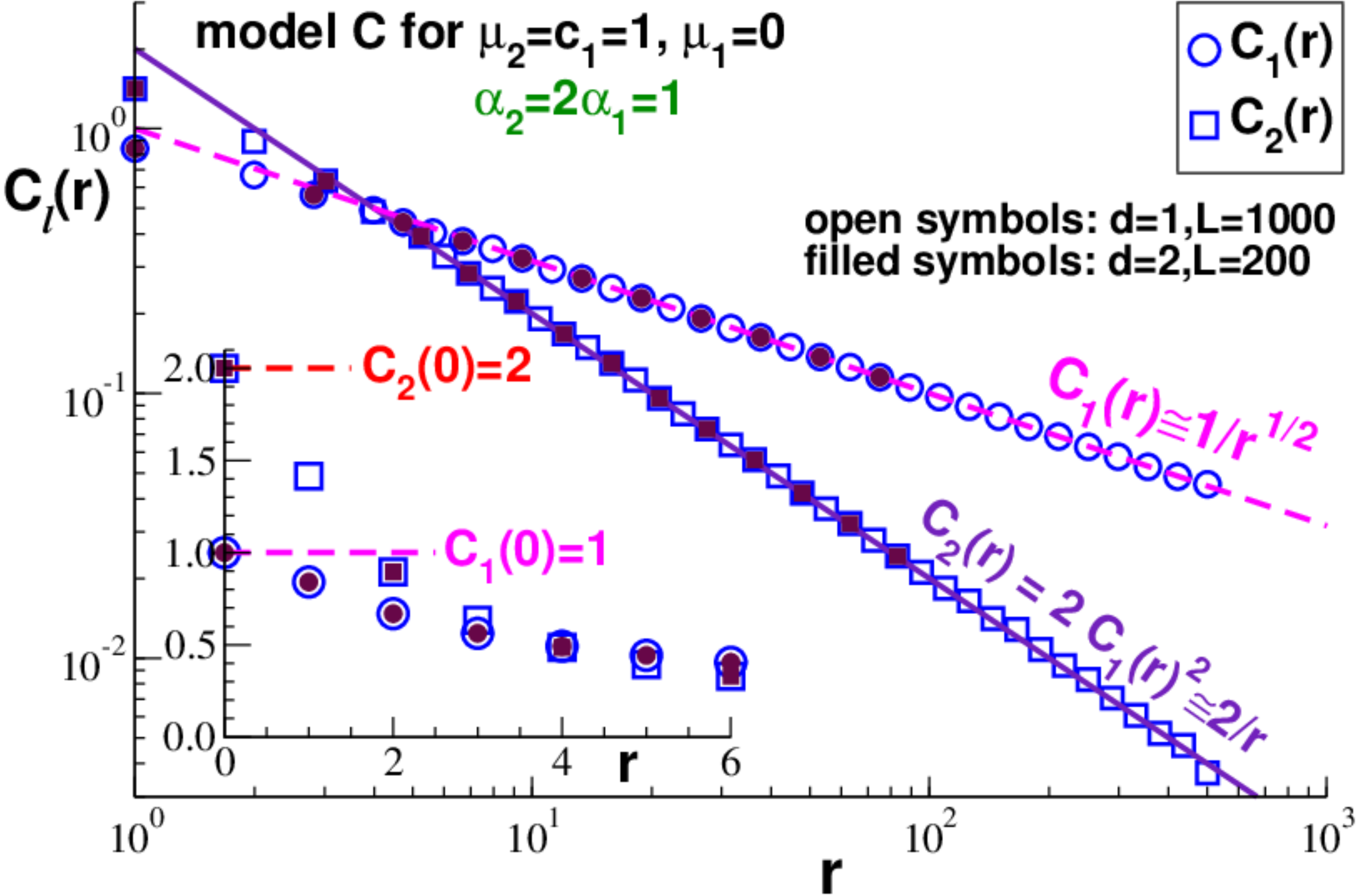}}}
\caption{$C_1(\rvec)$ and $C_2(\rvec)$ for model C with $\alpha_1=1/2$, $\mu_1=0$, $\mu_2=c_1=1$ and $\xi=1$.
The open symbols have been obtained for $d=1$ and $L=1000$, the filled symbols for $d=2$ and $L=200$.
Main: Double-logarithmic representation for logarithmically binned data. 
As emphasized by the solid line $\alpha_2=2\alpha_1=1$.
Inset: Linear representation for small $r$.
}
\label{fig_Cr_C}
\end{figure}
 
That the procedure works can be seen in Fig.~\ref{fig_Cr_C} for model C in $d=1$ and $d=2$ (filled symbols).  
We have set $\alpha_1 = 1/2$, $\mu_1=0$ and $\mu_2=c_1=1$. 
The data for $C_1(\rvec)$ and $C_2(\rvec)$ are obtained by averaging over $\Nc=10^4$ independent configurations. 
Due to Eq.~(\ref{eq_Wick3}) 
we have $C_2(\rvec) \simeq c_2/r^{\alpha_2}$ with $\alpha_2 = 2 \alpha_1 =1$
as shown by the solid line in the main panel.
The spatial dimension only plays a role for small and finite $r \approx 1$ 
as may be seen from the inset of Fig.~\ref{fig_Cr_C}. 
This leads to a weak $d$-dependence of integrals dominated by the lower integration bound.

\section{Alternative quenched field}
\label{app_X2}

An interesting alternative quenched field is given by
the ``local covariance" 
$v_{c\rvec} \equiv V \Eop^k \delta \sigma_{ck\rvec} \delta \sigma_{ck}$
between the local and total fluctuations $\delta \sigma_{ck\rvec}$ and 
$\delta \sigma_{ck} = \Eop^{\rvec} \delta \sigma_{ck\rvec}$.
Note that $v_c = \Eop^{\rvec} v_{c\rvec}$ holds since $\Eop^k$ and $\Eop^{\rvec}$ commute.
Importantly, for many physical systems $v_{c\rvec}$ corresponds to a local modulus,
e.g., the local stress-fluctuation contribution to an elastic modulus \cite{Lutsko89}. 
Using $\delta v_{c\rvec} = v_{c\rvec} -v$ we may write quite generally without 
any additional assumption
\begin{equation}
\Dnonerg = \Eop^{\rvec'} \Eop^{\rvec''} \underline{\Eop^{c} \delta v_{c\rvec'} \delta v_{c\rvec''}}
= \Eop^{\rvec} C[v_{\rvec}](\rvec)
\label{eq_Dnonerg_locelast}
\end{equation}
where $C[v_{\rvec}](\rvec)$ stands for the average of the underlined term over all pairs of 
microcells $\rvec'$ and $\rvec''=\rvec'+\rvec$.\footnote{Being
an $\rvec$-average over the pair correlation function of the covariance field, Eq.~(\ref{eq_Dnonerg_locelast})
is in fact a compact reformulation of the integral over the four-point correlation function
mentioned in Sec.~\ref{V_correl}. While it is easier numerically to deal with pair correlation 
functions, it is in general challenging to get a phenomenological or complete analytical
understanding of the scaling of $C[v_{\rvec}](\rvec)$. 
This is, however, possible for shear stresses in viscoelastic fluids
(including supercooled liquids and equilibrium amorphous systems)
where $v_{\rvec}$ is related to the local elastic shear modulus  \cite{spmP4}.}
As a consequence, a slow $V$-decrease of $\Snonerg$ with $\gamext < 1/2$ must arise if 
$C[v_{\rvec}](\rvec)$ is long-ranged. 
One may model the field $v_{\rvec}$ by means of a spatially correlated variable $\fr$,
i.e. using the notations of Sec.~\ref{spm_intro} we have $v = \mu_1$, 
$\Snonerg = \Delta_1$ and the correlation function $C_1(\rvec)$ corresponds to $C[v_{\rvec}](\rvec)$. 
This yields a good alternative fit of the shear-stress data discussed in Sec.~\ref{sec_simu} 
using model D with $\alpha_1=2$, $\mu_1=v=17.1$, $\mu_2=293$ and $\xi=3.3$.
To discriminate between both modeling approaches a numerical comparison of correlation functions 
of different $k$-averaged quenched fields is warranted \cite{spmP4}.
%


\end{document}